\documentclass[letterpaper]{article}
\usepackage[preprint]{aaai2027}
\nocopyright
\usepackage[hyphens]{url}
\usepackage{graphicx}
\usepackage{natbib}
\usepackage{caption}
\usepackage{subcaption}
\usepackage{booktabs}
\usepackage{amsmath}
\usepackage{amssymb}
\usepackage{xcolor}
\usepackage{enumitem}
\usepackage{array}
\usepackage{placeins}
\usepackage{tikz}
\usetikzlibrary{arrows.meta,positioning,fit,calc,shapes.geometric}
\definecolor{BAGTABlue}{HTML}{275D8C}
\definecolor{BAGTATeal}{HTML}{21867A}
\definecolor{BAGTAGold}{HTML}{C9861A}
\definecolor{BAGTARed}{HTML}{A33B3B}
\definecolor{BAGTAGray}{HTML}{5F6B76}
\newcolumntype{L}[1]{>{\raggedright\arraybackslash}p{#1}}

\title{GreenPassport: Request-Level Carbon Accounting for Cross-Border AI Inference}

\author{Rui Lu}
\affiliations{The Hong Kong Polytechnic University\\
ruilu@polyu.edu.hk}

\newcommand{\gco}{gCO$_2$e}
\newcommand{\kwh}{kWh}

\begin{document}

\maketitle

\begin{abstract}
AI inference often crosses regional boundaries as prompts travel to remote data centers and generated tokens return to users. Regional averages cannot represent the resulting differences in serving hardware, electricity, and network delivery. Request-level accounting needs a common boundary for the service, serving site, route, local comparator, uncertainty, and data provenance.
GreenPassport Carbon Accounting (GPCA) associates these inputs with each request. It estimates serving and route carbon, then selects a reporting level from the available documentation. Our public-data implementation covers data-center instances, accelerators, model families, electricity mixes, routes, and cloud-region carbon intensity.
Against six accounting baselines and four energy-prediction baselines, GPCA reduced median absolute percentage error by 56.3\% and median absolute error by 15.5\% relative to EcoLogits under the aligned accelerator-energy boundary. It produced zero rule overstatement in the deterministic conformance tests. In the buyer case, the clean-electricity CN-West scenario produced $0.0148$ gCO$_2$e /request, 88\% below the local service at $0.1220$ gCO$_2$e /request.
\end{abstract}

\section{Introduction}

AI inference is becoming a global service as LLMs enter applications and enterprise workflows. This demand requires accelerator clusters, stable electricity, and data-center capacity, but these resources are unevenly distributed across regions. As a result, inference may be consumed in one region and served in another. Customers send prompts via APIs or clouds, while generated tokens may return from distant sites with different electricity mixes, carbon intensities, and accelerator infrastructures.

\begin{figure*}[t]
    \centering
    \includegraphics[width=0.85\textwidth]{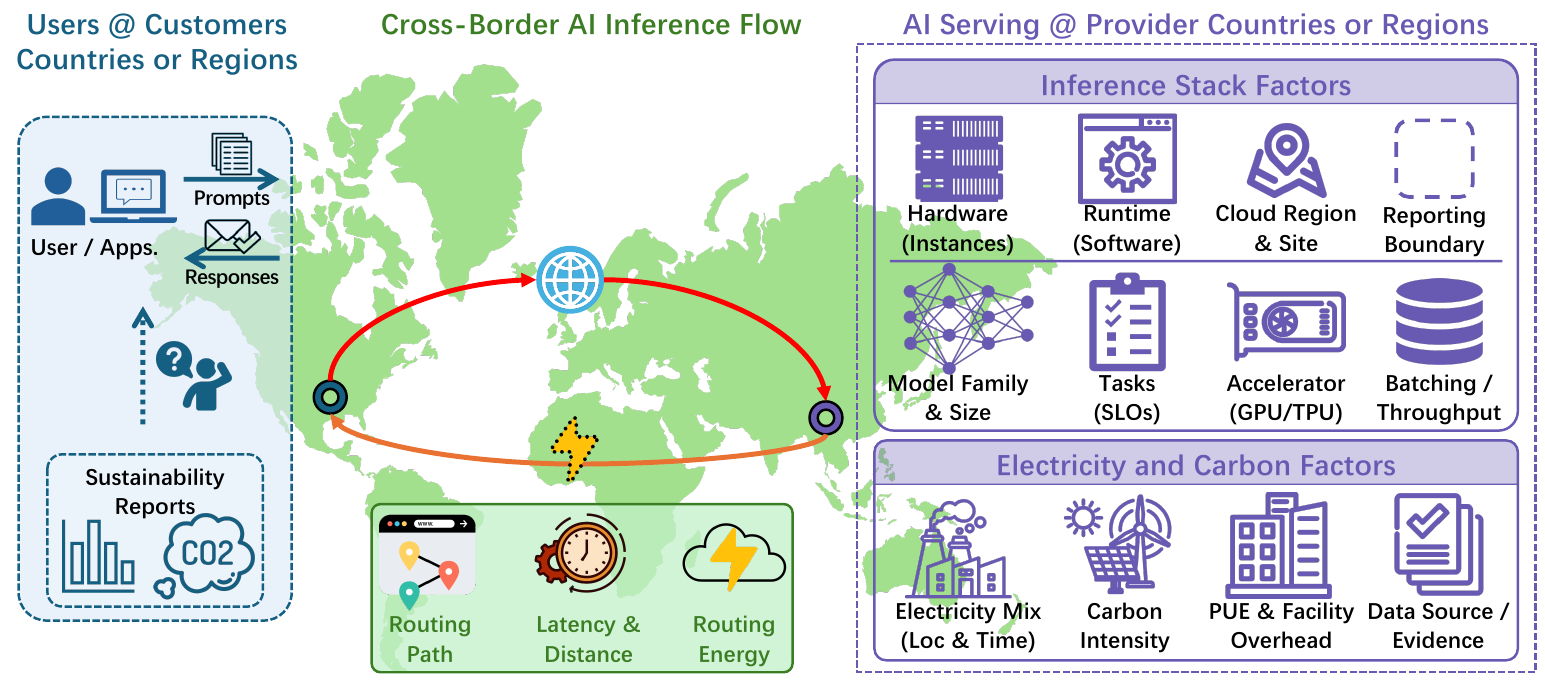}
    \caption{
    Cross-border AI inference flow. Customers send prompts from one region and receive responses from a distant provider. Request carbon depends on the serving stack, electricity and carbon factors, route, and documentation available for reporting.
    }\label{fig:cross-border-ai-inference}
\end{figure*}

Cross-border inference can be attractive as some regions offer cheaper energy, more available capacity, stronger grid interconnection, or easier data-center deployment. For example, in Fig.~\ref{fig:cross-border-ai-inference}, a US customer may purchase inference from a CN provider when local serving capacity is expensive. The operational footprint of this token service comes from the remote serving environment and delivery route. A request-level report needs the serving location, electricity source, hardware, route, and source documentation.

Data-center electricity demand has become an infrastructure constraint~\cite{iea2023datacentres,iea2025energyai}. Hardware, runtime, cloud region, and system boundary all affect ML carbon estimates~\cite{strubell2019energy,patterson2021carbon,henderson2020systematic,dodge2022measuring,luccioni2023bloom}. For inference, model family, task, accelerator, batching, and serving stack further change request energy~\cite{samsi2023wordswatts,elsworth2025google,wattcounts2026,mlenergy2026}. Carbon-aware operation adds location- and time-varying grid signals~\cite{radovanovic2022carbon,electricitymaps2026,watttime2026}, while market-based accounting adds procurement and anti-double-counting rules~\cite{ghgprotocol2015,energytag2022}. Comparing an imported service with a local deployment requires combining its serving energy with delivery carbon under the same request definition.

Request-level accounting must reconcile fragmented data. Hardware vendors publish accelerator and cloud-instance specifications, model reports give model sizes, and some providers publish service-level averages. Per-request energy, production routing, fleet composition, and serving sites are less often disclosed. Regional grid data also describe annual conditions more often than site conditions at request time. GPCA labels each input as measured, disclosed, estimated, or missing.

Carbon also varies among requests served for the same customer region. Model family, accelerator, batching, serving stack, route, request time, and electricity mix all affect the result. A regional annual average can hide these differences and misrank services. Cleaner electricity can accompany less efficient hardware, and a longer delivery route can offset part of the serving-site benefit. Each reported value includes its configuration and time basis.

Reporting eligibility depends on the input basis. Annual grid data, provider telemetry, hourly carbon data, route measurements, and certificates support different output levels. GPCA propagates their uncertainty through the serving and route terms before comparison with a named local alternative.

\textbf{GPCA} (\textbf{G}reen\textbf{P}assport \textbf{C}arbon \textbf{A}ccounting) combines request-level carbon estimation with source-aware reporting for cross-border AI inference. Each \textit{GreenPassport} records how a request is served and which reporting level its inputs support. GPCA computes request and token carbon from IT energy, facility overhead, serving-site carbon intensity, route bytes, and network energy. The paper makes three contributions.
\begin{itemize}[leftmargin=1.4em,itemsep=1pt,topsep=2pt]
    \item GPCA formulates cross-border carbon accounting at the request level and connects serving, electricity, route, uncertainty, and comparator inputs in one passport.
    \item Its accounting model separates measurement from fitted estimation, while its reporting rules map source quality and comparison stability to output levels.
    \item Evaluation on measured-energy data and public scenarios quantifies prediction error, configuration screening, regional sensitivity, and delivery effects.
\end{itemize}

\section{Background and Problem Setting}
\label{sec:background}

\subsection{Cross-Border Inference and Carbon Accounting}

An online inference request may execute in a remote data center before its generated output returns to the customer. Its operational footprint combines the serving environment, local power supply, and bidirectional delivery route.
The basic carbon identity remains energy multiplied by carbon intensity, with each term tied to a request. Compute energy comes from the model, hardware, runtime, batching, and output length. Facility overhead depends on how the serving site is operated. Carbon intensity depends on the site's location, time, electricity mix, and procurement treatment. Route energy is often small for short text but can become material for cross-region delivery or large payloads.
This setting requires the request definition, serving location, model, hardware, electricity basis, and delivery path. GPCA reports these inputs together with their uncertainty and provenance.

\subsection{Request-Level Accounting Setting}
GPCA reports the operational carbon of one inference request. A GreenPassport organizes the prompt, generated output, route payload, and serving context for this transaction.

Different parties observe different inputs. The customer knows the consumption region and may see an API name. The provider may know the model, accelerator, route, and telemetry but disclose only a subset. Independent verification can establish a higher reporting level.

One serving region can host many models, accelerators, instances, batching policies, and serving stacks. GPCA evaluates each service configuration within its serving region. Public annual grid data support estimates and scenarios. Provider telemetry, hourly carbon data, route measurements, and clean-power documents enable finer-grained output levels.

Two services are comparable when they serve the declared workload under compatible model availability, memory capacity, data-transfer policy, and latency or SLA class. Their carbon difference uses the same request definition and output length. Feasibility is checked before carbon calculation because a low regional factor can accompany higher serving energy, and an efficient accelerator may lack capacity for the selected model.

\section{Design of GreenPassport Carbon Accounting Model (GPCA)}
\label{sec:accounting_model}

\noindent\textbf{Request energy.}
Let request $r$ have input bytes $b_r^{\mathrm{in}}$ and output length $t_r^{\mathrm{out}}$. In a fully observed deployment, GPCA computes IT energy from serving-stack telemetry.
\[
E_{\mathrm{it}}(r,m,h,k)=\int_{t\in r} P_{\mathrm{it}}(t,m,h,k)\,dt ,
\]
where $P_{\mathrm{it}}$ is the serving-stack power attributable to the request. GPCA stores telemetry and public-data estimates under separate input modes.

\noindent\textbf{Shared-serving attribution.}
The telemetry identity also requires an attribution rule when requests share accelerators through continuous batching, shared KV cache, speculative decoding, host overhead, or multi-tenant serving. A measured passport states how batch or serving-window energy is assigned to requests. GPCA's default rule uses separate prefill and decode token-time after subtracting an idle baseline. Appendix~\ref{app:attribution_rule} gives the full rule and discusses its bias. A declared proxy, such as output-token or GPU-time allocation, maps to a lower reporting level.

\noindent\textbf{Calibrated public-data estimate.}
The public-data estimator represents model scale, output length, batching, accelerator family, device count, and mixture-of-experts (MoE) execution. We use the log-linear form
\begin{equation}
\begin{aligned}
\log \widehat E_{\mathrm{it}}
={}&\theta_0+\alpha\log A_m+\gamma\log t_r^{\mathrm{out}}
+\delta\log \bar b_k\\
&+\nu\log n_h+\mu\mathbf 1_{\mathrm{MoE}}+\eta_h ,
\end{aligned}
\nonumber
\end{equation}
where $A_m$ is active parameter count, $t_r^{\mathrm{out}}$ is mean output length for the workload, $\bar b_k$ is observed mean batch size, $n_h$ is accelerator count, and $\eta_h$ is an accelerator-family effect. Fitting on the ML.ENERGY calibration split gives $\alpha=0.616$, $\gamma=0.999$, $\delta=-0.536$, $\nu=0.556$, and $\mu=0.986$. The coefficients are frozen before exact-model-ID-grouped testing.
When batching or accelerator-specific effects are unavailable, GPCA uses a wider datasheet-based scenario and lists the missing variables. Closed services retain their disclosed service boundary. Appendix~Tables~\ref{tab:appendix_public_defaults}, \ref{tab:model_hierarchy}, and~\ref{tab:worked_request} give the defaults, input hierarchy, and a worked request.

A GreenPassport defines an energy-input set
\begin{equation}
\mathcal U_E(x)=
\begin{cases}
\{E_{\mathrm{it}}(x)\}, & \text{measured telemetry},\\
\{\widehat E_{\mathrm{it}}(x,\vartheta):\vartheta\in\Theta\},
& \text{public scenario},
\end{cases}
\nonumber
\end{equation}
where $\vartheta=(\theta_0,\alpha,\gamma,\delta,\nu,\mu,\eta_h)$ and $\Theta$ contains coefficient and residual uncertainty plus declared scenario bounds for missing inputs. The validation residual interval is distinct from route, PUE, and carbon-intensity bounds. Output levels use the combined uncertainty set.

\noindent\textbf{Uncertainty taxonomy.}
GPCA distinguishes three uncertainty categories. \emph{Measurement uncertainty} belongs to an observed power or electricity trace and its request-allocation procedure. \emph{Model residual uncertainty} comes from frozen-predictor errors on calibration data. The experiment uses a 90\% absolute log-residual envelope. \emph{Scenario or epistemic bounds} cover unobserved batching, PUE, site carbon intensity, route intensity, network electricity, and disclosure fields. The robust comparator combines the applicable sets and reports the limiting category.

Let $\rho_k$ be the PUE or facility multiplier for instance class $k$ and serving site $s$, and let $\widetilde E_{\mathrm{it}}$ denote either measured $E_{\mathrm{it}}$ or the public-data estimate $\widehat E_{\mathrm{it}}$. Facility energy equals IT energy multiplied by $\rho_k$. Multiplying facility energy in kWh by serving-site carbon intensity gives
\begin{equation}
C_{\mathrm{site}}(r,m,h,k,s)=\widetilde E_{\mathrm{it}}(r,m,h,k)\rho_k I_s/1000,
\nonumber
\end{equation}
where $I_s$ is serving-site carbon intensity in \gco/\kwh{} and energy is in Wh.

\begin{figure}[t]
\centering
\includegraphics[width=\linewidth]{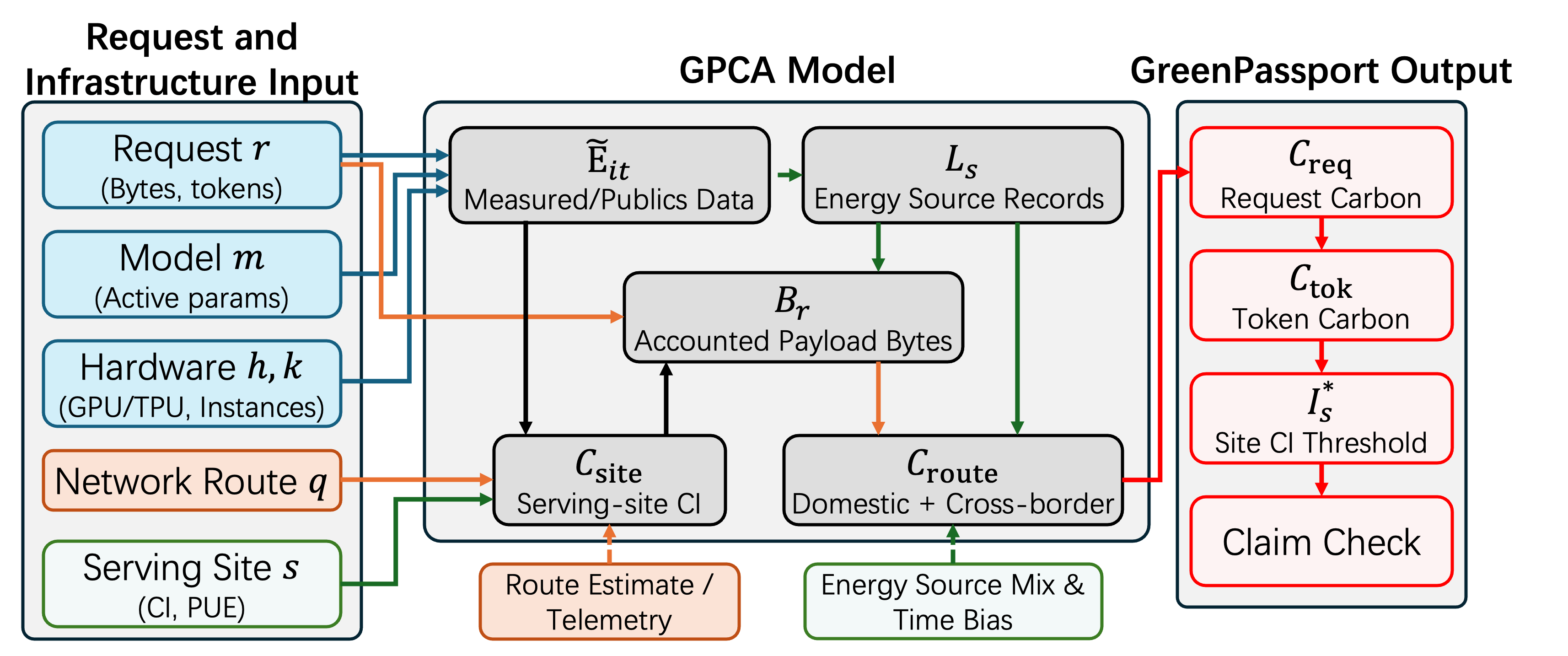}
\caption{Accounting components of GPCA.}
\label{fig:accounting_components}
\end{figure}

\noindent\textbf{Electricity-source input.}
For a serving site $s$, let $\mathcal G=\{\mathrm{coal},\mathrm{gas/oil},\mathrm{nuclear},\mathrm{hydro},\mathrm{wind},\mathrm{solar},\mathrm{bio/other}\}$ denote the reported generation-source set, and let $\pi_{s,g}$ be the annual generation share of source $g$. A source-resolved passport stores
\begin{equation}
L_s=\left(I_s,\{\pi_{s,g}\}_{g\in\mathcal G},\tau_s,\chi_s\right),
\nonumber
\end{equation}
where $\tau_s$ is the reporting tier and $\chi_s$ identifies a national, regional, annual, hourly, or certificate-backed basis. When source-specific emission factors $\kappa_{s,g}$ are available, the site term can be decomposed as
\begin{subequations}
\begin{equation}
C_{\mathrm{site}}^{(g)}(r,m,h,k,s)
=\widetilde E_{\mathrm{it}}(r,m,h,k)\rho_k\cdot\pi_{s,g}\kappa_{s,g}/1000.
\nonumber
\end{equation}
\begin{equation}
C_{\mathrm{site}}
=\sum_{g\in\mathcal G}C_{\mathrm{site}}^{(g)}.
\nonumber
\end{equation}
\end{subequations}
The public implementation uses independently verified annual regional carbon intensity $I_s$ for the carbon value and reports $\{\pi_{s,g}\}$ beside it.
Fig.~\ref{fig:accounting_components} shows how the request, infrastructure, site, and route inputs produce the GPCA outputs.

\noindent\textbf{Route energy.}
The route boundary is bidirectional. Let $\eta$ be response bytes per output token and $\omega$ be protocol overhead. The accounted payload combines prompt and response bytes with protocol overhead.
\begin{equation}
B_r=(b_r^{\mathrm{in}}+\eta t_r^{\mathrm{out}})(1+\omega).
\nonumber
\end{equation}
For a measured path, route $q$ is a set of segments $\{j\}$ with segment energy intensity $\epsilon_j$ and segment carbon intensity $I_j$. The route carbon term is
\begin{equation}
C_{\mathrm{route}}(r,q)=
\sum_{j\in q}\frac{B_r\epsilon_j I_j}{1000}.
\nonumber
\end{equation}
The prototype represents public route data as an interval set $\mathcal U_q$ over $\epsilon_j$ and $I_j$. Local service sets the cross-border segment to zero. For imported service, measured carrier or provider telemetry can replace the route estimate.

\noindent\textbf{Request and token carbon.}
The request-level and output-token-level values are
\begin{subequations}
\begin{equation}
C_{\mathrm{req}}(r,m,h,k,s,q)
=C_{\mathrm{site}}(r,m,h,k,s)+C_{\mathrm{route}}(r,q).
\nonumber
\end{equation}
\begin{equation}
C_{\mathrm{tok}}=C_{\mathrm{req}}/t_r^{\mathrm{out}}.
\nonumber
\end{equation}
\end{subequations}
The request value is primary. Token-normalized values help compare output lengths, but they can hide fixed prompt, routing, and service overheads.

\noindent\textbf{Memory feasibility.}
For model family $m$ and instance $k$, GPCA computes a coarse memory feasibility check.
\begin{equation}
G_{\min}(m,k)=
\left\lceil
\frac{\theta_m \lambda}{M_k \gamma_{\mathrm{mem}}}
\right\rceil ,
\nonumber
\end{equation}
where $\theta_m$ is total parameters, $\lambda$ is bytes per served parameter, $M_k$ is memory per accelerator, and $\gamma_{\mathrm{mem}}$ is a memory headroom factor. This feasibility check excludes single-instance configurations whose accelerator memory cannot host the selected model.

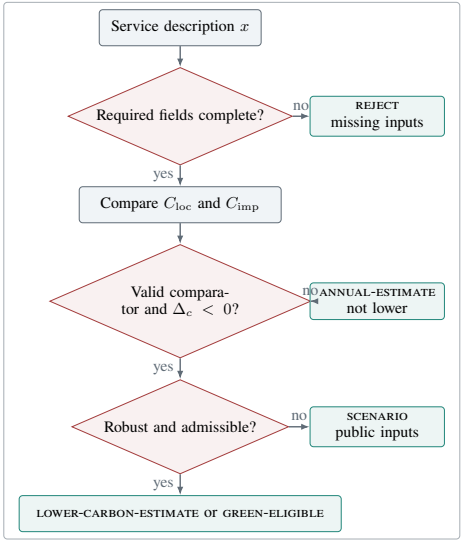
\begin{figure}[t]
\centering
\resizebox{0.725\columnwidth}{!}{%
\begin{tikzpicture}[
    node distance=0.42cm and 0.34cm,
    proc/.style={draw=BAGTAGray, rounded corners=2pt, align=center, minimum height=0.72cm, text width=3.0cm, font=\small, fill=BAGTABlue!7},
    gate/.style={diamond, aspect=2.3, draw=BAGTARed, align=center, inner sep=1pt, text width=3.60cm, font=\small, fill=BAGTARed!7},
    result/.style={draw=BAGTATeal, rounded corners=2pt, align=center, minimum height=0.72cm, text width=2.45cm, font=\small, fill=BAGTATeal!8},
    arr/.style={-{Latex[length=1.7mm]}, thick, draw=BAGTAGray},
    lab/.style={font=\small, text=BAGTAGray, align=center}
]
\node[proc] (service) {Service description $x$};
\node[gate, below=of service] (fields) {Required fields complete?};
\node[result, right=of fields] (reject) {\textsc{reject}\\missing inputs};
\node[proc, below=of fields, text width=3.8cm] (compare) {Compare $C_{\mathrm{loc}}$ and $C_{\mathrm{imp}}$};
\node[gate, below=of compare] (gap) {Valid comparator and $\Delta_c<0$?};
\node[result, anchor=east] (annual) at (reject.east |- gap) {\textsc{annual-estimate}\\not lower};
\node[gate, below=of gap] (adm) {Robust and admissible?};
\node[result, anchor=east] (scenario) at (reject.east |- adm) {\textsc{scenario}\\public inputs};
\node[result, below=of adm, text width=6.2cm] (label) {\textsc{lower-carbon-estimate} or \textsc{green-eligible}};

\draw[arr] (service) -- (fields);
\draw[arr] (fields) -- node[lab, left] {yes} (compare);
\draw[arr] (fields) -- node[lab, above] {no} (reject);
\draw[arr] (compare) -- (gap);
\draw[arr] (gap) -- node[lab, left] {yes} (adm);
\draw[arr] (gap) -- node[lab, above] {no} (annual);
\draw[arr] (adm) -- node[lab, left] {yes} (label);
\draw[arr] (adm) -- node[lab, above] {no} (scenario);

\node[draw=BAGTAGray!55, rounded corners=2pt, fit=(service)(fields)(gap)(adm)(label)(reject)(annual)(scenario),
      inner xsep=0.30cm, inner ysep=0.14cm] {};
\end{tikzpicture}
}
\caption{GreenPassport reporting workflow.}
\label{fig:protocol_workflow}
\end{figure}

\noindent\textbf{Reporting check.}
GPCA assigns a lower-carbon or green label only after comparing the request with a matched local service. For customer region $u$, define the comparator gap as
\begin{equation}
\Delta_c(r,m,h,k,s,q,u)=
\begin{aligned}[t]
&C_{\mathrm{req}}(r,m,h,k,s,q)\\
&{}-C_{\mathrm{req}}(r,m,h,k,u,q_{\mathrm{local}}).
\end{aligned}
\nonumber
\end{equation}
The reporting check evaluates the site input, route input, and uncertainty against the requested label.
For robust lower-carbon comparisons, GPCA uses the stronger condition
\begin{equation}
\sup_{z\in\mathcal U_{\mathrm{imp}}} C_{\mathrm{imp}}(z)
<
\inf_{z\in\mathcal U_{\mathrm{loc}}} C_{\mathrm{loc}}(z),
\nonumber
\end{equation}
where the uncertainty sets combine energy, route, facility, and carbon-intensity inputs. Public-data rows outside this robust condition receive an annual-estimate or scenario label.

\section{GreenPassport Reporting Standard}
\label{sec:reporting_standard}
Section~\ref{sec:accounting_model} computes request carbon, token carbon, and uncertainty from the request, service configuration, site, and route. The reporting standard adds a source basis, local comparator, and output level. Following model cards, data statements, and datasheets, GPCA makes required fields and provenance explicit \cite{mitchell2019modelcards,bender2018datastatements,gebru2021datasheets}. Fig.~\ref{fig:protocol_workflow} summarizes the sequence.
Let
\[
x=(r,m,h,k,s,q,a,\mathcal D)
\]
denote one service description. It contains request/workload $r$, model or service $m$, hardware $h$, instance class $k$, serving site $s$, delivery route $q$, input basis $a$, and supporting documents $\mathcal D$. With comparator record $d_c$, the required accounting fields are
\[
\mathcal R=\{r,m,s,q,a,\mathcal D,d_c\}\cup \{h,k\}.
\]
The standard rejects incomplete entries and compares each remaining service against two local comparators. Here, $c_{\mathrm{same}}$ is the same workload/model/hardware where locally available, and $c_{\mathrm{best}}$ is the lowest-carbon feasible local hardware in the declared catalog. A comparator is valid under an operational predicate $\Omega(x,c)$ covering local model availability, customer-side data constraints, latency or SLA class, and any capacity or price condition used in the customer report. An invalid comparator record yields an annual estimate. Missing records or failed service constraints yield rejection.

\noindent\textbf{Request-level selector.}
The passport enables GPCA to select among a defined set of per-request deployment candidates. For customer region $u$, let $\mathcal F(r,u)$ contain candidate tuples $(m,h,k,s,q,a,\mathcal D)$ that satisfy the required fields, memory feasibility, route boundary, and operational predicate for request $r$. GPCA selects
\begin{equation}
x^\star(r,u)=
\arg\min_{\substack{(m,h,k,s,q,a,\mathcal D)\\\in\mathcal F(r,u)}}
C_{\mathrm{req}}(r,m,h,k,s,q),
\label{eq:gpca_selector}
\end{equation}
where $x^\star$ stores the selected model, accelerator, instance, serving site, route, input basis, and documents. Admissibility constrains the minimization, and the rule below assigns the selected row an annual-estimate, lower-carbon-estimate, or green-eligible label.

The two local comparators answer different buyer questions. The same-configuration comparator isolates the location and route effect for an otherwise matched service. The best-local comparator asks whether the imported candidate improves on any feasible local alternative in the declared catalog. GPCA reports both gaps because an imported service may improve on a matched local configuration while remaining above a more efficient local deployment. The operational predicate is evaluated before either gap enters the label rule, so infeasible or policy-incompatible candidates do not define the reported minimum.

Let $\mathrm{adm}(x,\ell)$ require a valid comparator and inputs for label level $\ell$. Annual national or regional data support annual estimates. Official accelerator and instance specifications support scenarios. Provider energy disclosures apply to their stated service boundary. Drawing on the GHG Protocol and EnergyTag, GPCA requires timestamped workload data, deliverability, residual-mix treatment, and no double counting for hourly or certificate-backed green labels \cite{ghgprotocol2015,energytag2022}. Define missing fields $M(x)=\mathcal R\setminus\mathrm{fields}(x)$ and comparator gap $\Delta_c(x)=C_{\mathrm{imp}}(x)-C_{\mathrm{loc}}(x,c)$. A lower-carbon label also requires the robust inequality in Section~\ref{sec:accounting_model} when public uncertainty sets are non-singletons. Let $O_x$ and $V_c$ denote service feasibility and comparator validity. Evaluate the cases in order. The standard returns
\begin{equation*}
y(x,c)=
\begin{cases}
\textsc{reject}, & M(x)\neq\emptyset\vee\neg O_x,\\
\textsc{annual-estimate}, & \neg V_c\vee\Delta_c(x)\ge 0,\\
\textsc{scenario}, &
\substack{V_c\wedge\Delta_c(x)<0\wedge\\ \neg\mathrm{adm}(x,\textsc{lower})},\\
\substack{\textsc{lower-carbon-}\\\textsc{estimate}}, &
\substack{V_c\wedge\Delta_c(x)<0\wedge\\
\mathrm{adm}(x,\textsc{lower})\wedge\\
\neg\mathrm{adm}(x,\textsc{green})},\\
\textsc{green-eligible}, &
\substack{V_c\wedge\Delta_c(x)<0\wedge\\ \mathrm{adm}(x,\textsc{green})}.
\end{cases}
\end{equation*}
A machine-readable GreenPassport stores the selected deployment and its accounting result. An independent verifier assigns the effective reporting level from the supporting documents. Appendix~\ref{app:implementation_details} gives the schema and input defaults.

\noindent\textbf{Disclosure checks.}
The verifier checks timestamps, model--hardware feasibility, workload boundaries, allocation rules, route inputs, and clean-electricity documents. Appendix~\ref{app:reporting_algorithm} gives the full decision table and consistency checks.

\section{Evaluation and Implementation}
\label{sec:evaluation}
\subsection{Implementation}
We implement GPCA as a public-data pipeline. 
The platform stage reads regional grid data, Google Cloud region-carbon rows \cite{googlecloudregion2024}, accelerator and instance specifications, TPU slices, open-model facts, closed-service energy disclosures, route defaults, and workload profiles.
The accounting stage combines these data with source mix, uncertainty, attribution, comparator, and reporting settings to produce GreenPassport rows. 
The pipeline writes the calculated values and reporting decisions to the passport schema.
The evaluation uses configuration files and processed public data with recorded source provenance.

\subsection{Evaluation Setup}
\label{subsec:eval_setup}

The evaluation combined public energy measurements, cross-border accounting scenarios, and deterministic reporting-rule tests. It examined predictive validity, output stability under uncertainty, tool boundaries, and component removal. Twenty fixtures per method tested conformance to the reporting rules. Global sensitivity used 50,000 deterministic samples from the declared input ranges. The reported frequencies are proportions under the sampling distribution. A fixture overstated its result when its requested label exceeded the level returned by the reporting rule. The test compared requested and returned labels for each case and aggregated the mismatch rate. Component-omission regret used the lowest-carbon feasible and admissible candidate under GPCA as the reference.

\noindent\textbf{Open measured-energy dataset and split.}
We used public aggregate measurements from the ML.ENERGY leaderboard covering LLM chat, GPQA, and code fill-in-the-middle tasks on H100 and B200 accelerators \cite{chung2025mlenergy}. We split the data by exact model identifier, excluding every validation identifier from fitting. The target was accelerator energy per response and excluded PUE, network, and end-user energy.

\noindent\textbf{Default candidates.}
The cross-border stress test compared US-Middle local service with US-East, US-West, CN-East, and CN-West for a medium-assistant workload. The buyer case added four public GCP U.S. regions. CN-East was an annual-grid negative control, and CN-West was a hypothetical clean-electricity scenario.

All candidates used the same request boundary, including prompt payload, generated output, facility multiplier, and bidirectional route accounting. The frozen estimator supplied serving energy for configurations covered by the calibration fields. Missing configuration inputs invoked the declared scenario range. Site carbon used the annual, regional, or scenario intensity, and route carbon used domestic or cross-border route classes. Every comparison retained the same functional unit of grams of CO$_2$e per request. Token carbon divided this total by generated output length.

Global sensitivity varied serving configuration, electricity, and delivery inputs jointly. PUE ranged from 1.10 to 1.50. Local carbon intensity ranged from 250 to 500 \gco/\kwh, CN-East from 450 to 650, and the CN-West clean scenario from 5 to 150. Route energy intensity ranged from 0.006 to 0.60 kWh/GB, with network carbon intensity between 300 and 600 \gco/\kwh. Prompt payload ranged from 10 KB to 1 MB, output length from 100 to 4,000 tokens, and mean batch size from 8 to 32. The energy interval used a multiplicative residual factor of $1.918$. These scenario bounds describe the input combinations tested in the regional comparison. Appendix~Table~\ref{tab:sensitivity_ranges} gives the remaining coefficient ranges.

\begin{figure}[t]
\centering
\setlength{\abovecaptionskip}{3pt}
\begin{minipage}[b]{0.49\linewidth}
\centering
\includegraphics[width=\linewidth]{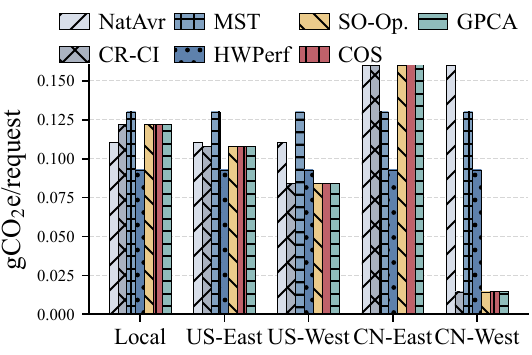}
\captionof{figure}{Request carbon ($\downarrow$).}
\label{fig:eval4_request_carbon}
\end{minipage}\hfill
\begin{minipage}[b]{0.49\linewidth}
\centering
\includegraphics[width=\linewidth]{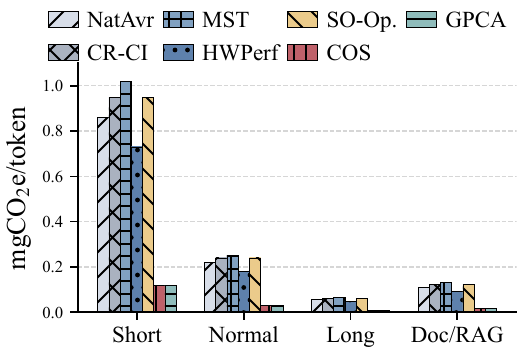}
\captionof{figure}{Token carbon ($\downarrow$).}
\label{fig:eval5_token_carbon}
\end{minipage}
\end{figure}

\noindent\textbf{Baselines.}
We compared six baselines that isolate the accounting and selection controls combined by GPCA.
\begin{itemize}[leftmargin=*,nosep]
\item National-Average Carbon Intensity (\textit{NatAvr}). Applies fixed request energy and annual national grid intensity \cite{lannelongue2021green}.
\item Cloud-Region Carbon Intensity (\textit{CR-CI}). Replaces the national factor with a regional or cloud-region factor while retaining fixed request energy \cite{anthony2020carbontracker}.
\item Model-Size and Token Proxy (\textit{MST}). Predicts request energy from model size and output length \cite{luccioni2024power}.
\item Hardware Performance-per-Watt Proxy (\textit{HWPerf}). Uses nominal accelerator performance per watt as its energy proxy \cite{strubell2019energy}.
\item Site-Only Operational Accounting (\textit{SO-Op.}). Combines IT energy, PUE, and serving-site carbon intensity within the data-center boundary \cite{luccioni2023bloom}.
\item Carbon-Only Selector (\textit{COS}). Chooses the candidate with the lowest derived operational-carbon point estimate \cite{acun2023carbon}.
\end{itemize}
All methods received the same candidate set, workload profiles, and public carbon inputs. Each method used the fields and decision controls listed above.

\noindent\textbf{Predictors and tool boundaries.}
To evaluate predictive validity, we compared a training-set median, model- and accelerator-based proxies, the calibrated GPCA proxy, and EcoLogits \cite{rince2025ecologits,ecologits082}. We normalized the EcoLogits GPU-energy component by the observed device count and batch size and compared it on the accelerator-energy-per-response target. Appendix~Table~\ref{tab:tool_boundary} documents the boundaries of CodeCarbon, CarbonTracker, EcoLogits, and GPCA \cite{lacoste2019quantifying,anthony2020carbontracker}.

\noindent\textbf{Operational predicate.}
The buyer-facing selector requires model availability, memory feasibility, route disclosure, data-transfer permission, a compatible latency class, and source documentation. Deployment recommendations additionally require comparable price, capacity, and private-SLA data.

\noindent\textbf{Metrics.}
We report median absolute percentage error, median absolute error, and Spearman rank correlation for energy prediction. Top-1 agreement and measured regret evaluate configuration selection, and validation-interval coverage evaluates uncertainty. The reporting metrics cover comparator gap, conditional output, route contribution, and overstatement under component removal. Appendix~\ref{app:evaluation_metrics} gives the definitions.

\begin{figure}[t]
\centering
\setlength{\abovecaptionskip}{3pt}
\begin{subfigure}[t]{0.49\linewidth}
\centering
\includegraphics[width=\linewidth]{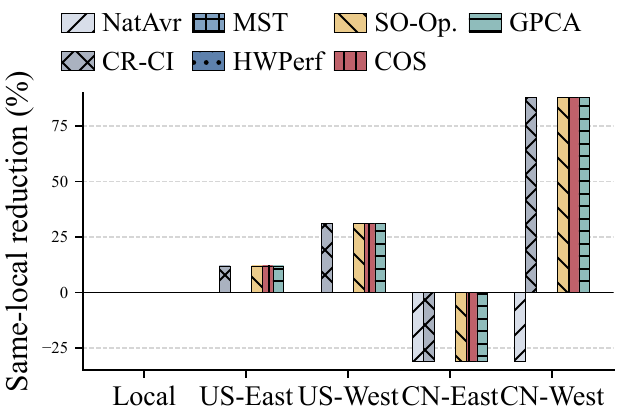}
\caption{$R_{\mathrm{same}}$ ($\uparrow$)}
\label{fig:eval6_same_local}
\end{subfigure}\hfill
\begin{subfigure}[t]{0.49\linewidth}
\centering
\includegraphics[width=\linewidth]{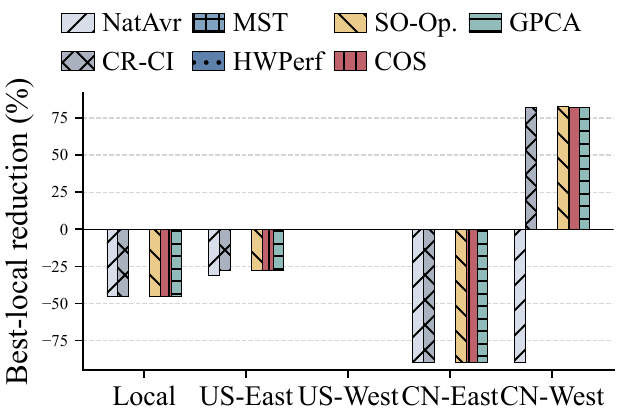}
\caption{$R_{\mathrm{best}}$ ($\uparrow$)}
\label{fig:eval6_best_local}
\end{subfigure}
\caption{Comparator reductions ($\uparrow$).}
\label{fig:eval6_comparator_reduction}
\end{figure}

\begin{figure*}[t]
\centering
\begin{minipage}[b]{0.21\textwidth}
\centering
\includegraphics[width=\linewidth]{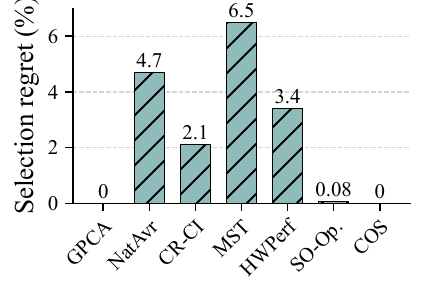}
\captionof{figure}{Component-omission regret fixture ($\downarrow$).}
\label{fig:eval7_selection_regret}
\end{minipage}\hfill
\begin{minipage}[b]{0.59\textwidth}
\centering
\setlength{\abovecaptionskip}{1pt}
\setlength{\belowcaptionskip}{1pt}
\begin{subfigure}[t]{0.333\linewidth}
\centering
\includegraphics[width=\linewidth]{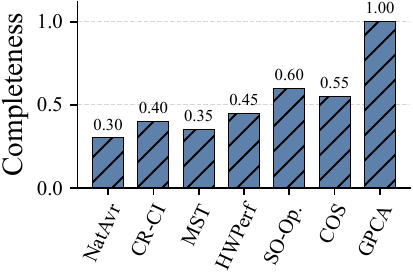}
\caption{Fields ($\uparrow$)}
\end{subfigure}\hspace{-0.012\linewidth}
\begin{subfigure}[t]{0.333\linewidth}
\centering
\includegraphics[width=\linewidth]{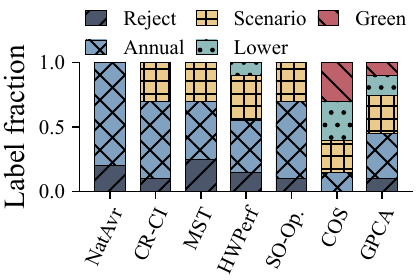}
\caption{Labels}
\end{subfigure}\hspace{-0.012\linewidth}
\begin{subfigure}[t]{0.333\linewidth}
\centering
\includegraphics[width=\linewidth]{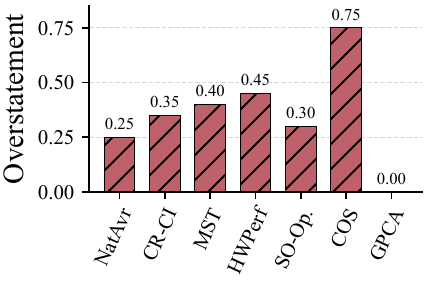}
\caption{Overstatement ($\downarrow$)}
\end{subfigure}
\caption{Reporting-rule conformance fixtures.}
\label{fig:eval8_reporting_quality}
\end{minipage}\hfill
\begin{minipage}[b]{0.19\textwidth}
\centering
\includegraphics[width=\linewidth]{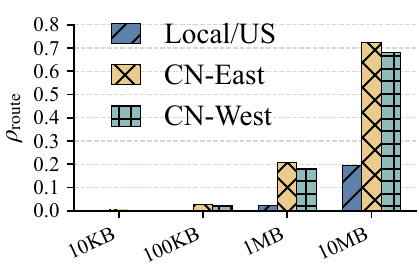}
\captionof{figure}{Route contribution ($\downarrow$).}
\label{fig:eval9_route_contribution}
\end{minipage}
\end{figure*}

\subsection{Results}
\label{subsec:overall}

\noindent\textbf{Prediction and ranking.}
The calibrated GPCA proxy achieved the lowest median APE at 23.0\% and median MAE at 0.0169 Wh per response (Table~\ref{tab:energy_validation}). Median APE was 18.9\% for chat, 28.4\% for GPQA, and 31.5\% for code fill-in-the-middle. EcoLogits had the highest rank correlation at 0.964, compared with 0.934 for GPCA. Both methods achieved 66.7\% top-1 agreement and zero median measured selection regret. The nominal 90\% GPCA residual interval covered 84.0\% of the validation data. Regional comparisons combined this interval with scenario bounds for the remaining inputs.

\begin{table}[h]
\centering
\setlength{\belowcaptionskip}{2pt}
\setlength{\abovecaptionskip}{2pt}
\caption{Energy-prediction validation.}
\label{tab:energy_validation}
\small
\resizebox{\columnwidth}{!}{%
\begin{tabular}{@{}lrrrrr@{}}
\toprule
Estimator & Median APE & Median MAE & Spearman $\rho$ & Top-1 agreement & Median regret \\
\midrule
Constant & 212.3\% & 0.1413 Wh & 0.000 & 0.0\% & 629.9\% \\
Model proxy & 88.0\% & 0.0969 Wh & 0.748 & 0.0\% & 120.4\% \\
Accelerator proxy & 56.5\% & 0.0463 Wh & 0.596 & 0.0\% & 120.4\% \\
EcoLogits (aligned) & 52.6\% & 0.0200 Wh & \textbf{0.964} & \textbf{66.7\%} & \textbf{0.0\%} \\ \hline
GPCA & \textbf{23.0\%} & \textbf{0.0169 Wh} & 0.934 & \textbf{66.7\%} & \textbf{0.0\%} \\
\bottomrule
\end{tabular}
}
\end{table}
 
\noindent\textbf{Regional and request accounting.}
Across the sensitivity samples, CN-East was above the local comparator in 82.1\% of cases and overlapped it in 17.9\%. CN-West was lower in 42.5\%, overlapped in 45.1\%, and was higher in 12.4\%. Figs.~\ref{fig:eval4_request_carbon} and~\ref{fig:eval5_token_carbon} show how the model, accelerator, instance, route, and source mix changed request and token carbon.

\begin{table}[t]
\centering
\setlength{\belowcaptionskip}{2pt}
\setlength{\abovecaptionskip}{2pt}
\caption{Buyer comparison and GPCA results.}
\label{tab:eval_buyer_carbon_gap}
\small
\setlength{\tabcolsep}{3pt}
\resizebox{\linewidth}{!}{
\begin{tabular}{@{}lrrrrl@{}}
\toprule
& \multicolumn{2}{c}{Carbon (\gco/request)} & \multicolumn{2}{c}{Reduction (\%)} & \\
\cmidrule(lr){2-3}\cmidrule(lr){4-5}
Candidate & Route & Total & Same local & Best local & GPCA result \\
\midrule
Local   & 0.00004 & 0.1220 & 0   & -434 & Local comparator \\
US-East & 0.00003 & 0.1078 & 12  & -372 & Not best-local \\
US-West & 0.00002 & 0.0843 & 31  & -269 & Not best-local \\
GCP-Oregon & 0.00004 & 0.0229 & 81 & 0 & Best-local comparator \\
GCP-N. Virginia & 0.00004 & 0.0931 & 24 & -307 & Interval overlap \\
GCP-Iowa & 0.00004 & 0.1189 & 3 & -420 & Not lower \\
GCP-S. Carolina & 0.00004 & 0.1659 & -36 & -626 & Not lower \\
CN-East & 0.00038 & 0.1604 & -31 & -602 & Not lower \\
CN-West & 0.00038 & 0.0148 & 88  & 35  & Conditional scenario \\
\bottomrule
\end{tabular}
}
\end{table}
 
\noindent\textbf{Selection and reporting.}
A candidate could outperform the same local configuration while exceeding the best feasible local service (Fig.~\ref{fig:eval6_comparator_reduction}). Removing selection controls increased regret (Fig.~\ref{fig:eval7_selection_regret}). In the deterministic tests, field completeness ranged from 0.30 to 0.60 for the baselines and was 0.55 for the carbon-only selector, compared with 1.00 for GPCA. Rule overstatement ranged from 0.25 to 0.75 for the baselines and was zero for GPCA (Fig.~\ref{fig:eval8_reporting_quality}). Route carbon was small for short text and increased with payload, reaching material shares for larger requests (Fig.~\ref{fig:eval9_route_contribution}).

\subsection{Case Study}
\label{subsec:case_study}

We studied a US-Middle buyer procuring medium-assistant tokens for an interactive LLM API. The candidate set contained nine services across the buyer's local region, other U.S. regions, and China. Selection minimized request carbon subject to model availability, memory feasibility, data-transfer policy, latency compatibility, and source documentation. The inputs combined ML.ENERGY aggregate measurements~\cite{chung2025mlenergy}, public accelerator and cloud specifications, electricity data~\cite{owidember2026,emberus2026,googlecloudregion2024}, and literature-bounded route intensity~\cite{aslan2018electricity,coroama2018internet}. Section~\ref{subsec:interpretation} examines the effects of the accounting components on this decision.

The reference request contained a 10 KB prompt and 500 output tokens. With four bytes per output token and 15\% protocol overhead, its bidirectional payload was 13.8 KB. For the CN-West scenario, the reference IT-energy estimate was 0.24 Wh. A PUE of 1.2 and site carbon intensity of 50 \gco/\kwh{} gave $0.0144$ \gco{} of site carbon. Adding the route estimate of $0.00038$ \gco{} yielded the reported total.

Table~\ref{tab:eval_buyer_carbon_gap} compares the nine candidates. US-East and US-West had lower request carbon than the matched local service, but both exceeded GCP Oregon. Thus, moving the matched configuration to another region improved its carbon estimate while a different local deployment offered a larger reduction. Appendix~Fig.~\ref{fig:eval10_buyer_funnel} shows the comparison steps.

At the reference point, GPCA selected the CN-West clean scenario at $0.0148$ \gco/request. This estimate was 88\% below US-Middle local service and 35\% below GCP Oregon, the best U.S. comparator. GCP Oregon was 81\% below local service, whereas CN-East and GCP S. Carolina were 31\% and 36\% higher. Delivery contributed $0.00038$ \gco/request, or 2.6\% of the CN-West total.

\subsection{Interpretation}
\label{subsec:interpretation}

GPCA reduced absolute prediction error relative to EcoLogits, while EcoLogits achieved higher rank correlation. Both methods had the same top-1 agreement and zero median selection regret. Validation excluded the exact model identifiers used for fitting and tested transfer within the represented H100 and B200 configurations.

The buyer comparison shows why serving location and hardware must be considered together. US-East and US-West improved on the matched local configuration, but GCP Oregon remained lower. The memory check further restricted which configurations entered this comparison. Route energy introduced a separate dependence on payload, increasing the delivery share for long requests and document/RAG workloads. These effects determine the lowest-carbon feasible service for a given request.

We removed one accounting component at a time while retaining the workload data, public inputs, and carbon coefficients. NoRoute removed delivery carbon, and NoInstanceGate admitted configurations that failed the memory check. SameLocalOnly compared an imported service only with the matched local configuration. PointEstimate applied the reporting rule to point values, and NoSourceGate removed the source and time-basis requirements. These variants tested which parts of the calculation changed the carbon total, candidate set, or reporting level.

NoRoute had the largest effect on requests with large payloads. Delivery accounted for 0.25\% of the short-request total and 1.96\% of the normal-request total, increasing to 16.45\% for long requests and 66.27\% for document/RAG. Removing this term therefore changed the total most for document delivery. NoInstanceGate changed the configurations available to the selector. The large model fit the H100 and B200 entries, while the MoE workload fit the B200 entry. Feasible medium-model estimates ranged from 0.11 to 0.43 \gco{} across the represented accelerators.

SameLocalOnly changed the reference against which a reduction was reported. In the buyer case, the CN-West reduction decreased from 88\% against the matched local service to 35\% against the best feasible local service. NoSourceGate assigned a lower-carbon or green label to 85\% of its entries. The label fraction decreased as uncertainty widened and reached zero when required inputs were missing. Appendix~\ref{app:ablation_study} provides the workload curves and component settings.

\section{Related Work}
\label{app:related_work}

\noindent\textbf{Carbon measurement and inference energy.}
CodeCarbon measures or estimates local hardware energy \cite{lacoste2019quantifying}, and CarbonTracker predicts the footprint of a running training job \cite{anthony2020carbontracker}. EcoLogits estimates the usage and embodied impacts of remote generative-AI APIs \cite{rince2025ecologits}. Inference measurements further show variation across model size, task length, batching, accelerator generation, memory behavior, and serving implementation \cite{samsi2023wordswatts,elsworth2025google,wattcounts2026,mlenergy2026,wang2025storellm,lu2025taws}. 

\noindent\textbf{Serving systems, electricity, and delivery.}
Carbon-aware datacenter systems use location- and time-varying grid intensity for workload placement and scheduling \cite{radovanovic2022carbon,acun2023carbon}. Electricity Maps documents consumption-based average signals \cite{electricitymaps2026}, while WattTime provides marginal location and time signals \cite{watttime2026}.  Network studies relate delivery energy to traffic volume, path, utilization, and system boundary \cite{aslan2018electricity,coroama2018internet,tabaeiaghdaei2022carbon}.

\noindent\textbf{Accounting boundary and documentation.}
GHG Protocol Scope 2 defines location- and market-based accounting, residual-mix treatment, and double-counting guidance \cite{ghgprotocol2015}. EnergyTag specifies granular certificate schemes \cite{energytag2022}. Model cards \cite{mitchell2019modelcards}, data statements \cite{bender2018datastatements}, and datasheets \cite{gebru2021datasheets} structure the documentation of models and datasets. GreenPassport applies these documentation principles to request-carbon accounting and local-service comparisons.

\section{Conclusion}

GPCA combines serving-energy estimation and route accounting to compute the operational carbon of cross-border inference requests. GreenPassport compares each service with local alternatives and assigns a reporting level using input provenance and uncertainty. Evaluation showed lower absolute prediction error than EcoLogits on public measured-energy data and zero rule overstatement in deterministic tests. The buyer case quantified how comparator choice changed the reported reduction, while the workload analysis showed how delivery contributed to request carbon. Together, these results connect request-carbon estimates to service comparisons under explicit workload, deployment, and electricity conditions.

\clearpage
\bibliography{references}

@misc{owidember2026,
  author       = {{Our World in Data} and {Ember}},
  title        = {Carbon intensity of electricity and electricity generation by source},
  year         = {2026},
  howpublished = {\url{https://ourworldindata.org/grapher/carbon-intensity-electricity.csv}},

}

@misc{emberus2026,
  author       = {{Ember}},
  title        = {United States Yearly Electricity Data},
  year         = {2026},
  howpublished = {\url{https://storage.googleapis.com/emb-prod-bkt-publicdata/public-downloads/us_yearly_full_release_long_format.csv}},

}

@misc{ghgprotocol2015,
  author       = {{World Resources Institute} and {World Business Council for Sustainable Development}},
  title        = {GHG Protocol Scope 2 Guidance},
  year         = {2015},
  howpublished = {\url{https://ghgprotocol.org/scope_2_guidance}}
}

@misc{energytag2022,
  author       = {{EnergyTag}},
  title        = {Granular Certificate Scheme Standard, Version 2},
  year         = {2024},
  howpublished = {\url{https://energytag.org/wp-content/uploads/2024/12/EnergyTag_Granular-Certificate-Scheme-Standard-V2.pdf}},
  note         = {Versioned granular-certificate scheme standard}
}

@inproceedings{mitchell2019modelcards,
  author    = {Mitchell, Margaret and Wu, Simone and Zaldivar, Andrew and Barnes, Parker and Vasserman, Lucy and Hutchinson, Ben and Spitzer, Elena and Raji, Inioluwa Deborah and Gebru, Timnit},
  title     = {Model Cards for Model Reporting},
  booktitle = {Proceedings of the Conference on Fairness, Accountability, and Transparency},
  pages     = {220--229},
  year      = {2019},
  doi       = {10.1145/3287560.3287596}
}

@article{bender2018datastatements,
  author  = {Bender, Emily M. and Friedman, Batya},
  title   = {Data Statements for Natural Language Processing: Toward Mitigating System Bias and Enabling Better Science},
  journal = {Transactions of the Association for Computational Linguistics},
  year    = {2018},
  volume  = {6},
  pages   = {587--604},
  doi     = {10.1162/tacl_a_00041}
}

@article{gebru2021datasheets,
  author  = {Gebru, Timnit and Morgenstern, Jamie and Vecchione, Briana and Vaughan, Jennifer Wortman and Wallach, Hanna and Daum{\'e} III, Hal and Crawford, Kate},
  title   = {Datasheets for Datasets},
  journal = {Communications of the ACM},
  year    = {2021},
  volume  = {64},
  number  = {12},
  pages   = {86--92},
  doi     = {10.1145/3458723}
}

@misc{iea2025energyai,
  author       = {{International Energy Agency}},
  title        = {Energy and AI},
  year         = {2025},
  howpublished = {\url{https://www.iea.org/reports/energy-and-ai}}
}

@misc{iea2023datacentres,
  author       = {{International Energy Agency}},
  title        = {Data Centres and Data Transmission Networks},
  year         = {2023},
  howpublished = {\url{https://www.iea.org/energy-system/digitalisation/data-centres-and-data-transmission-networks}}
}

@misc{electricitymaps2026,
  author       = {{Electricity Maps}},
  title        = {Electricity Maps Methodology},
  year         = {2026},
  howpublished = {\url{https://www.electricitymaps.com/data/methodology}},

}

@misc{watttime2026,
  author       = {{WattTime}},
  title        = {Methodology and Validation},
  year         = {2026},
  howpublished = {\url{https://watttime.org/data-science/methodology-validation/}},

}

@article{patterson2021carbon,
  author  = {Patterson, David and Gonzalez, Joseph and Le, Quoc and Liang, Chen and Munguia, Lluis-Miquel and Rothchild, Daniel and So, David and Texier, Maud and Dean, Jeff},
  title   = {Carbon Emissions and Large Neural Network Training},
  journal = {arXiv preprint arXiv:2104.10350},
  year    = {2021}
}

@inproceedings{strubell2019energy,
  author    = {Strubell, Emma and Ganesh, Ananya and McCallum, Andrew},
  title     = {Energy and Policy Considerations for Deep Learning in NLP},
  booktitle = {Proceedings of the 57th Annual Meeting of the Association for Computational Linguistics},
  pages     = {3645--3650},
  year      = {2019},
  doi       = {10.18653/v1/P19-1355}
}

@article{henderson2020systematic,
  author  = {Henderson, Peter and Hu, Jieru and Romoff, Joshua and Brunskill, Emma and Jurafsky, Dan and Pineau, Joelle},
  title   = {Towards the Systematic Reporting of the Energy and Carbon Footprints of Machine Learning},
  journal = {Journal of Machine Learning Research},
  year    = {2020},
  volume  = {21},
  number  = {248},
  pages   = {1--43}
}

@inproceedings{dodge2022measuring,
  author    = {Dodge, Jesse and Prewitt, Taylor and Tachet des Combes, Remi and Odmark, Erika and Schwartz, Roy and Strubell, Emma and Luccioni, Alexandra Sasha and Smith, Noah A. and DeCario, Nicole and Buchanan, Will},
  title     = {Measuring the Carbon Intensity of AI in Cloud Instances},
  booktitle = {Proceedings of the 2022 ACM Conference on Fairness, Accountability, and Transparency},
  year      = {2022},
  doi       = {10.1145/3531146.3533234}
}

@article{lacoste2019quantifying,
  author  = {Lacoste, Alexandre and Luccioni, Alexandra and Schmidt, Victor and Dandres, Thomas},
  title   = {Quantifying the Carbon Emissions of Machine Learning},
  journal = {arXiv preprint arXiv:1910.09700},
  year    = {2019}
}

@inproceedings{anthony2020carbontracker,
  author    = {Anthony, Lasse F. Wolff and Kanding, Benjamin and Selvan, Raghavendra},
  title     = {Carbontracker: Tracking and Predicting the Carbon Footprint of Training Deep Learning Models},
  booktitle = {Proceedings of the 2020 ICML Workshop on Challenges in Deploying and Monitoring Machine Learning Systems},
  year      = {2020}
}

@article{lannelongue2021green,
  author  = {Lannelongue, Lo{\"i}c and Grealey, Jason and Inouye, Michael},
  title   = {Green Algorithms: Quantifying the Carbon Footprint of Computation},
  journal = {Advanced Science},
  year    = {2021},
  volume  = {8},
  number  = {12},
  pages   = {2100707},
  doi     = {10.1002/advs.202100707}
}

@article{luccioni2023bloom,
  author    = {Luccioni, Alexandra Sasha and Viguier, Sylvain and Ligozat, Anne-Laure},
  title     = {Estimating the Carbon Footprint of BLOOM, a 176B Parameter Language Model},
  journal   = {Journal of Machine Learning Research},
  year      = {2023},
  volume    = {24},
  number    = {253},
  pages     = {1--15}
}

@inproceedings{luccioni2024power,
  author    = {Luccioni, Alexandra Sasha and Jernite, Yacine and Strubell, Emma},
  title     = {Power Hungry Processing: Watts Driving the Cost of AI Deployment?},
  booktitle = {Proceedings of the 2024 ACM Conference on Fairness, Accountability, and Transparency},
  year      = {2024},
  pages     = {85--99}
}

@inproceedings{acun2023carbon,
  author    = {Acun, Bilge and Lee, Benjamin and Kazhamiaka, Fiodar and Maeng, Kiwan and Gupta, Udit and Chakkaravarthy, Manoj and Brooks, David and Wu, Carole-Jean},
  title     = {Carbon Explorer: A Holistic Framework for Designing Carbon Aware Datacenters},
  booktitle = {Proceedings of the 28th ACM International Conference on Architectural Support for Programming Languages and Operating Systems},
  year      = {2023},
  pages     = {118--132}
}

@article{elsworth2025google,
  author  = {Elsworth, Cooper and Huang, Keguo and Patterson, David and Schneider, Ian and Sedivy, Robert and Goodman, Savannah and Townsend, Ben and Ranganathan, Parthasarathy and Dean, Jeff and Vahdat, Amin and Gomes, Ben and Manyika, James},
  title   = {Measuring the Environmental Impact of Delivering AI at Google Scale},
  journal = {arXiv preprint arXiv:2508.15734},
  year    = {2025}
}

@article{samsi2023wordswatts,
  author  = {Samsi, Siddharth and Zhao, Dan and McDonald, Joseph and Li, Baolin and Michaleas, Adam and Jones, Michael and Bergeron, William and Kepner, Jeremy and Tiwari, Devesh and Gadepally, Vijay},
  title   = {From Words to Watts: Benchmarking the Energy Costs of Large Language Model Inference},
  journal = {arXiv preprint arXiv:2310.03003},
  year    = {2023}
}

@article{masanet2020recalibrating,
  author  = {Masanet, Eric and Shehabi, Arman and Lei, Nuoa and Smith, Sarah and Koomey, Jonathan},
  title   = {Recalibrating Global Data Center Energy-Use Estimates},
  journal = {Science},
  year    = {2020},
  volume  = {367},
  number  = {6481},
  pages   = {984--986},
  doi     = {10.1126/science.aba3758}
}

@article{radovanovic2022carbon,
  author  = {Radovanovi{\'c}, Ana and Koningstein, Ross and Schneider, Ian and Chen, Bokan and Duarte, Alexandre and Roy, Binz and Xiao, Diyue and Haridasan, Maya and Hung, Patrick and Care, Nick and Talukdar, Saurav and Mullen, Eric and Smith, Kendal and Cottman, MariEllen and Cirne, Walfredo},
  title   = {Carbon-Aware Computing for Datacenters},
  journal = {IEEE Transactions on Power Systems},
  year    = {2023},
  volume  = {38},
  number  = {2},
  pages   = {1270--1280},
  doi     = {10.1109/TPWRS.2022.3173250}
}

@article{aslan2018electricity,
  author  = {Aslan, Joshua and Mayers, Kieren and Koomey, Jonathan G. and France, Chris},
  title   = {Electricity Intensity of Internet Data Transmission: Untangling the Estimates},
  journal = {Journal of Industrial Ecology},
  year    = {2018},
  volume  = {22},
  number  = {4},
  pages   = {785--798},
  doi     = {10.1111/jiec.12630}
}

@article{coroama2018internet,
  author  = {Coroama, Vlad C. and Hilty, Lorenz M.},
  title   = {Assessing Internet Energy Intensity: A Review of Methods and Results},
  journal = {Environmental Impact Assessment Review},
  year    = {2014},
  volume  = {45},
  pages   = {63--68}
}

@article{tabaeiaghdaei2022carbon,
  author  = {Tabaeiaghdaei, Seyedali and Scherrer, Simon and Kwon, Jonghoon and Perrig, Adrian},
  title   = {Carbon-Intelligent Global Routing in Path-Aware Networks},
  journal = {arXiv preprint arXiv:2211.00347},
  year    = {2022}
}

@misc{nvidiaa100,
  author       = {{NVIDIA}},
  title        = {NVIDIA A100 Tensor Core GPU},
  year         = {2026},
  howpublished = {\url{https://www.nvidia.com/en-us/data-center/a100/}},

}

@misc{nvidiah100,
  author       = {{NVIDIA}},
  title        = {NVIDIA H100 Tensor Core GPU},
  year         = {2026},
  howpublished = {\url{https://www.nvidia.com/en-eu/data-center/h100/}},

}

@misc{nvidiah200,
  author       = {{NVIDIA}},
  title        = {NVIDIA H200 Tensor Core GPU},
  year         = {2026},
  howpublished = {\url{https://www.nvidia.com/en-us/data-center/h200/}},

}

@misc{nvidial40s,
  author       = {{NVIDIA}},
  title        = {NVIDIA L40S GPU},
  year         = {2026},
  howpublished = {\url{https://www.nvidia.com/en-us/data-center/l40s/}},

}

@misc{nvidiadgxb200,
  author       = {{NVIDIA}},
  title        = {NVIDIA DGX B200 Specifications},
  year         = {2026},
  howpublished = {\url{https://www.nvidia.com/en-us/data-center/dgx-b200/}},

}

@misc{amdmi300x,
  author       = {{Advanced Micro Devices}},
  title        = {AMD Instinct MI300X Accelerator},
  year         = {2026},
  howpublished = {\url{https://www.amd.com/en/products/accelerators/instinct/mi300/mi300x.html}},

}

@misc{awsp5,
  author       = {{Amazon Web Services}},
  title        = {Amazon EC2 P5 Instances},
  year         = {2026},
  howpublished = {\url{https://aws.amazon.com/ec2/instance-types/p5/}},

}

@misc{awsp6,
  author       = {{Amazon Web Services}},
  title        = {Amazon EC2 P6e UltraServers and P6 Instances},
  year         = {2026},
  howpublished = {\url{https://aws.amazon.com/ec2/instance-types/p6/}},

}

@misc{gcpa3,
  author       = {{Google Cloud}},
  title        = {Accelerator-Optimized Machine Family},
  year         = {2026},
  howpublished = {\url{https://cloud.google.com/compute/docs/accelerator-optimized-machines}},

}

@misc{googletpuv4,
  author       = {{Google Cloud}},
  title        = {TPU v4},
  year         = {2026},
  howpublished = {\url{https://cloud.google.com/tpu/docs/v4}},

}

@misc{googletpuv5p,
  author       = {{Google Cloud}},
  title        = {TPU v5p},
  year         = {2026},
  howpublished = {\url{https://cloud.google.com/tpu/docs/v5p}},

}

@misc{googletpuv6e,
  author       = {{Google Cloud}},
  title        = {TPU v6e},
  year         = {2026},
  howpublished = {\url{https://cloud.google.com/tpu/docs/v6e}},

}

@misc{azurendh100,
  author       = {{Microsoft Azure}},
  title        = {ND-H100-v5 Size Series},
  year         = {2026},
  howpublished = {\url{https://learn.microsoft.com/en-us/azure/virtual-machines/sizes/gpu-accelerated/ndh100v5-series}},

}

@misc{googlecloudregion2024,
  author       = {{Google Cloud}},
  title        = {Carbon Free Energy for Google Cloud Regions: 2024 Data},
  year         = {2026},
  howpublished = {\url{https://cloud.google.com/sustainability/region-carbon}},

}

@article{deepseekv3,
  author  = {{DeepSeek-AI}},
  title   = {DeepSeek-V3 Technical Report},
  journal = {arXiv preprint arXiv:2412.19437},
  year    = {2024}
}

@article{qwen25,
  author  = {{Qwen Team}},
  title   = {Qwen2.5 Technical Report},
  journal = {arXiv preprint arXiv:2412.15115},
  year    = {2024}
}

@misc{llama31modelcard,
  author       = {{Meta}},
  title        = {Llama 3.1 Model Card},
  year         = {2024},
  howpublished = {\url{https://github.com/meta-llama/llama-models/blob/main/models/llama3_1/MODEL_CARD.md}},
  note         = {Reports 8B, 70B, and 405B model sizes}
}

@misc{mixtral8x7b,
  author       = {{Mistral AI}},
  title        = {Mixtral 8x7B Model Card},
  year         = {2023},
  howpublished = {\url{https://docs.mistral.ai/models/model-cards/mixtral-8x7b-0-1}},
  note         = {Reports 47B total and 13B active parameters}
}

@misc{altman2025gentle,
  author       = {Altman, Sam},
  title        = {The Gentle Singularity},
  year         = {2025},
  howpublished = {\url{https://blog.samaltman.com/the-gentle-singularity}},
  note         = {Includes a public average ChatGPT query energy statement}
}

@article{wattcounts2026,
  author  = {Argerich, Mauricio Fadel and F{\"u}rst, Jonathan and Pati{\~n}o-Mart{\'i}nez, Marta},
  title   = {Watt Counts: Energy-Aware Benchmark for Sustainable LLM Inference on Heterogeneous GPU Architectures},
  journal = {arXiv preprint arXiv:2604.09048},
  year    = {2026}
}

@misc{mlperf51,
  author       = {{MLCommons}},
  title        = {MLPerf Inference v5.1 Benchmark Results},
  year         = {2025},
  howpublished = {\url{https://mlcommons.org/2025/09/mlperf-inference-v5-1-results/}},

}

@misc{mlenergy2026,
  author       = {Chung, Jae-Won},
  title        = {Diagnosing Inference Energy Consumption with the ML.ENERGY Leaderboard v3.0},
  year         = {2026},
  howpublished = {\url{https://ml.energy/blog/measurement/energy/diagnosing-inference-energy-consumption-with-the-mlenergy-leaderboard-v30/}},

}

@inproceedings{chung2025mlenergy,
  author    = {Chung, Jae-Won and Ma, Jeff J. and Wu, Ruofan and Liu, Jiachen and Kweon, Oh Jun and Xia, Yuxuan and Wu, Zhiyu and Chowdhury, Mosharaf},
  title     = {The {ML.ENERGY} Benchmark: Toward Automated Inference Energy Measurement and Optimization},
  booktitle = {Advances in Neural Information Processing Systems Datasets and Benchmarks Track},
  year      = {2025},
  url       = {https://github.com/ml-energy/leaderboard},

}

@article{rince2025ecologits,
  author  = {Rinc{\'e}, Samuel and Banse, Adrien},
  title   = {{EcoLogits}: Evaluating the Environmental Impacts of Generative AI},
  journal = {Journal of Open Source Software},
  year    = {2025},
  volume  = {10},
  number  = {111},
  pages   = {7471},
  doi     = {10.21105/joss.07471}
}

@misc{ecologits082,
  author       = {Rinc{\'e}, Samuel and Banse, Adrien and {EcoLogits contributors}},
  title        = {{EcoLogits} 0.8.2},
  year         = {2025},
  howpublished = {\url{https://pypi.org/project/ecologits/0.8.2/}},

}

@inproceedings{lepikhin2021gshard,
  author    = {Lepikhin, Dmitry and others},
  title     = {GShard: Scaling Giant Models with Conditional Computation and Automatic Sharding},
  booktitle = {International Conference on Learning Representations},
  year      = {2021}
}

@inproceedings{dao2022flashattention,
  author    = {Dao, Tri and Fu, Daniel Y. and Ermon, Stefano and Rudra, Atri and R{\'e}, Christopher},
  title     = {FlashAttention: Fast and Memory-Efficient Exact Attention with IO-Awareness},
  booktitle = {Advances in Neural Information Processing Systems},
  year      = {2022}
}

@inproceedings{leviathan2023speculative,
  author    = {Leviathan, Yaniv and Kalman, Matan and Matias, Yossi},
  title     = {Fast Inference from Transformers via Speculative Decoding},
  booktitle = {International Conference on Machine Learning},
  year      = {2023}
}

@inproceedings{dettmers2023qlora,
  author    = {Dettmers, Tim and Pagnoni, Artidoro and Holtzman, Ari and Zettlemoyer, Luke},
  title     = {QLoRA: Efficient Finetuning of Quantized LLMs},
  booktitle = {Advances in Neural Information Processing Systems},
  year      = {2023}
}

@inproceedings{dao2024flashattention2,
  author    = {Dao, Tri},
  title     = {FlashAttention-2: Faster Attention with Better Parallelism and Work Partitioning},
  booktitle = {International Conference on Learning Representations},
  year      = {2024}
}

@inproceedings{wang2025storellm,
  author = {Wang, Dan and Liu, Boan and Lu, Rui and Zhang, Zhaorui and Zhu, Shuntao},
  title = {{StoreLLM}: Energy Efficient Large Language Model Inference with Permanently Pre-stored Attention Matrices},
  booktitle = {Proceedings of the 16th ACM International Conference on Future and Sustainable Energy Systems},
  pages = {398--406},
  year = {2025},
  doi = {10.1145/3679240.3734604}
}

@inproceedings{lu2025taws,
  author = {Lu, Rui and Wang, Dan},
  title = {A Thermal-aware Workload Scheduler for High-performance {LLM} Inference in Cooling-regulated Datacenters},
  booktitle = {Proceedings of the 4th Workshop on Sustainable Computer Systems},
  year = {2025},
  doi = {10.1145/3757892.3757906}
}

\clearpage
\appendix
\numberwithin{figure}{section}
\numberwithin{table}{section}
\renewcommand{\theHfigure}{appendix.\thesection.\arabic{figure}}
\renewcommand{\theHtable}{appendix.\thesection.\arabic{table}}

\section{Public Inputs and Provenance}
\label{app:implementation_details}

This appendix gives the public inputs, predictor checks, reporting rules, and component ablations used in the evaluation.

\subsection{Model Hierarchy and Worked Request}
Appendix~Table~\ref{tab:model_hierarchy} separates measured quantities, calibrated public proxies, scenario inputs, uncertainty, and label decisions. Appendix~Table~\ref{tab:worked_request} applies that hierarchy to the reference request and gives its scenario-level result.

\begin{table*}[t]
\centering
\caption{Accounting and reporting hierarchy.}
\label{tab:model_hierarchy}
\scriptsize
\resizebox{\textwidth}{!}{
\begin{tabular}{@{}llllll@{}}
\toprule
Quantity & Definition & Observed mode & Public proxy & Uncertainty type & Used in \\
\midrule
$E_{\mathrm{it}}$ & Attributed serving energy & Power trace and allocation rule & None & Measurement & Site carbon \\
$\widehat E_{\mathrm{it}}$ & Request-energy predictor & Workload and configuration fields & Calibrated aggregate model & Residual/model & Screening and ranking \\
$\mathcal U_E$ & Energy input set & Meter uncertainty & Residual plus missing-input envelope & Statistical and epistemic & Robust comparison \\
$C_{\mathrm{site}}$ & Facility operational carbon & PUE and timestamped site CI & Public PUE and regional CI & Scenario/epistemic & Request carbon \\
$C_{\mathrm{route}}$ & Delivery carbon & Segment telemetry & Route-class intensity interval & Scenario/epistemic & Request carbon \\
$\ell^\star$ & Strongest admissible label & Verified provenance and comparator & Public input values & Conditional decision & Report output \\
\bottomrule
\end{tabular}}
\end{table*}
\begin{table*}[t]
\centering
\caption{Worked request accounting.}
\label{tab:worked_request}
\small
\resizebox{\textwidth}{!}{%
\begin{tabular}{@{}llll@{}}
\toprule
Stage & Input or rule & Result & Source status \\
\midrule
Workload & 10 KB prompt, 500 output tokens & 13.8 KB routed bytes with overhead & Declared request \\
IT energy & Reference estimator member & 0.24 Wh/request & Public scenario \\
Facility & PUE $1.2$, site CI $50$ gCO$_2$/kWh & $0.0144$ \gco/request & Unverified clean-site scenario \\
Route & $0.006/0.06$ kWh/GB domestic/cross-border & $0.00038$ \gco/request & Aggregate route-class proxy \\
Total & Site plus route & $0.01478$ \gco/request & Point estimate \\
Comparator & US-West point estimate $0.0843$ \gco/request & $-82.5\%$ point gap & Public comparator \\
Decision & Residual plus global epistemic envelope & Conditional regional result & Label varies across the envelope \\
\bottomrule
\end{tabular}
}
\end{table*}

\subsection{Public-Scenario Defaults}
Annual grid carbon intensities and generation mixes come from the open OWID/Ember datasets. The evaluation uses US-Middle/local, US-East, US-West, GCP Oregon, N. Virginia, Iowa, S. Carolina, CN-East, and CN-West candidates \cite{owidember2026,emberus2026}. Google Cloud 2024 region CFE and grid-carbon values are fetched from Google's machine-readable region-carbon dataset when a cloud-region row is needed \cite{googlecloudregion2024}. The route configuration defines the reference prompt size $b_0$, output length $t_0$, response-byte factor $\eta$, protocol overhead $\omega$, and domestic or CN-to-US route-energy intensities. Table~\ref{tab:appendix_public_defaults} lists the numerical defaults used by the public run. The route, overhead, and payload values are declared stress-test settings drawn from literature-informed ranges.

\begin{table}[h]
\centering
\caption{Public-scenario defaults and sweep values.}
\label{tab:appendix_public_defaults}
\scriptsize
\begin{tabular}{@{}L{0.58\linewidth}L{0.30\linewidth}@{}}
\toprule
Symbol notation & Value \\
\midrule
$E_0$ (legacy reference IT-energy anchor) & $0.24$ Wh/request \\
$(\alpha,\gamma,\delta)$ (fitted model, output, and batch coefficients) & $0.616$, $0.999$, $-0.536$ \\
$(\nu,\mu)$ (fitted GPU-count and MoE coefficients) & $(0.556,0.986)$ \\
$\eta_{\mathrm{H100}}$ (fitted H100 effect relative to B200) & $-0.276$ log-joules \\
Calibration residual envelope & $1.918\times$ symmetric multiplicative factor \\
$\xi_k$ (datasheet-fallback instance or host overhead) & $1.08$ \\
$\zeta_{\mathrm{MoE}}$ (datasheet-fallback MoE overhead) & $1.08$ \\
$\rho$ (reference PUE multiplier) & $1.20$ \\
$b_0$ (reference prompt payload) & $10$ KB \\
$t_0$ (reference output length) & $500$ tokens \\
$\eta$ (response bytes per output token) & $4$ bytes/token \\
$\omega$ (protocol overhead) & $15\%$ \\
$\epsilon_{\mathrm{dom}}$ (domestic route-energy intensity) & $0.006$ kWh/GB \\
$\epsilon_{\mathrm{CN\to US}}$ (CN-to-US route-energy intensity) & $0.06$ kWh/GB \\
$\lambda$ (bytes per served parameter) & $1$ byte/parameter \\
$\gamma_{\mathrm{mem}}$ (usable accelerator-memory share) & $0.75$ \\
$b_r^{\mathrm{in}}$ sweep (prompt-payload stress grid) & $10$ KB, $100$ KB, $1$ MB, $10$ MB \\
\bottomrule
\end{tabular}
\end{table}

\subsection{Global-Sensitivity Ranges}
Table~\ref{tab:sensitivity_ranges} gives the declared ranges used for the global sensitivity analysis in Section~\ref{subsec:overall}. Model and serving ranges surround the fitted ML.ENERGY coefficients. PUE, grid, and route ranges are screening envelopes informed by the public inputs and network-energy literature \cite{chung2025mlenergy,aslan2018electricity,coroama2018internet,tabaeiaghdaei2022carbon}. The analysis uses their declared endpoints as epistemic stress bounds.
\begin{table}[h]
\centering
\caption{Declared global-sensitivity ranges.}
\label{tab:sensitivity_ranges}
\scriptsize
\setlength{\tabcolsep}{2pt}
\begin{tabular}{@{}p{0.31\linewidth}p{0.28\linewidth}p{0.37\linewidth}@{}}
\toprule
Input & Range & Basis \\
\midrule
Model exponent $\alpha$ & $0.45$--$0.80$ & Around fitted $0.616$ \\
Output exponent $\gamma$ & $0.85$--$1.15$ & Around fitted $0.999$ \\
Batch exponent $\delta$ & $-0.75$--$-0.30$ & Around fitted $-0.536$ \\
Imported accelerator factor & $0.65$--$1.15$ & Broad H100/B200 scenario \\
Instance overhead & $1.00$--$1.20$ & Declared host envelope \\
PUE & $1.10$--$1.50$ & Facility scenario envelope \\
Local CI & $250$--$500$ gCO$_2$/kWh & Public-grid envelope \\
CN-East CI & $450$--$650$ gCO$_2$/kWh & Annual-grid envelope \\
CN-West scenario CI & $5$--$150$ gCO$_2$/kWh & Clean-scenario envelope \\
Route energy intensity & $0.006$--$0.60$ kWh/GB & Literature stress envelope \\
Route carbon intensity & $300$--$600$ gCO$_2$/kWh & Mixed-network envelope \\
Prompt payload & $10$ KB--$1$ MB & Text/RAG profiles \\
Output length & $100$--$4{,}000$ tokens & Serving profiles \\
Mean batch size & $8$--$32$ & ML.ENERGY support \\
Residual factor & $1.918\times$ & Calibration 90\% log residual \\
\bottomrule
\end{tabular}
\end{table}

\subsection{Hardware and Estimate Checks}
\begin{figure}[h]
\centering
\includegraphics[width=0.96\linewidth]{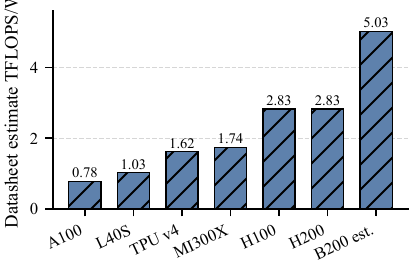}
\caption{AI-datacenter accelerator catalog used by the public-data estimate ($\uparrow$). Performance-per-watt values come from public specifications. TPU v4 uses Google Cloud's published mean measured chip power. B200 uses a DGX B200 system-power allocation because the source set lacks a directly comparable GPU TDP.}
\label{fig:app_a1_hardware_efficiency}
\end{figure}

The hardware layer represents seven AI-datacenter accelerator classes, including A100, L40S, H100, H200, Blackwell-class B200, AMD MI300X, and Google Cloud TPU v4. Public B200/GB200 system specifications provide the Blackwell-class entry. TPU v4 uses Google's published peak BF16 compute, HBM, and measured min/mean/max chip power. TPU v5p and TPU v6e remain specification entries because their public pages lack comparable chip-power values for TFLOPS/W conversion. The instance layer uses public accelerator instances and TPU slices from AWS, Google Cloud, and Azure \cite{nvidiaa100,nvidial40s,nvidiah100,nvidiah200,nvidiadgxb200,amdmi300x,googletpuv4,googletpuv5p,googletpuv6e,awsp5,awsp6,gcpa3,azurendh100}. Fig.~\ref{fig:app_a1_hardware_efficiency} summarizes the catalog.

Table~\ref{tab:appendix_public_defaults} lists the fitted ML.ENERGY coefficients and residual envelope. The public leaderboard identifies batching, accelerator family, active parameters, and output length as separate fields \cite{chung2025mlenergy}. Exact-model-ID-grouped validation measures their predictive value. Watt Counts and provider studies also show that model, hardware, workload, and serving configuration affect request energy \cite{wattcounts2026,elsworth2025google}. Workloads and accelerators outside the calibration range use the wider datasheet-based scenario.

\subsection{Model and Service Data}
The model layer separates open model families from closed provider services. DeepSeek, Qwen, Llama, and Mixtral enter accelerator-specific scenario matrices using public model-size information \cite{deepseekv3,qwen25,llama31modelcard,mixtral8x7b}. ChatGPT and Gemini enter through service-level disclosures. The inputs use Sam Altman's 0.34 Wh average ChatGPT query statement and Google's 0.24 Wh median Gemini Apps text-prompt measurement \cite{altman2025gentle,elsworth2025google}. Their public disclosures omit the serving fleet, route, and workload mix. Table~\ref{tab:model_layer} lists the per-family facts and reporting scope.

\begin{table}[h]
\centering
\caption{Public model and service inputs.}
\label{tab:model_layer}
\scriptsize
\resizebox{\linewidth}{!}{
\begin{tabular}{@{}L{0.19\linewidth}L{0.18\linewidth}L{0.17\linewidth}L{0.12\linewidth}L{0.22\linewidth}@{}}
\toprule
Family/service & Public model facts & Public energy fact & Accel. matrix & Input basis \\
\midrule
ChatGPT & N/A & 0.34 Wh average query & No & Service disclosure only \\
Gemini Apps text & N/A & 0.24 Wh median prompt & No & Production-stack median \\
DeepSeek V3/R1 & 671B total, 37B active & N/A & Yes & Scenario only \\
Qwen2.5-72B & 72B dense family & N/A & Yes & Scenario only \\
Qwen2.5-32B & 32B dense family & N/A & Yes & Scenario only \\
Llama 3.1-70B & 70B dense family & N/A & Yes & Scenario only \\
Llama 3.1-405B & 405B dense family & N/A & Yes & Large-model scenario \\
Mixtral-8x7B & 47B total, 13B active & N/A & Yes & MoE scenario \\
\bottomrule
\end{tabular}
}
\end{table}

\subsection{Schema, Provenance, and Generated Artifacts}
The machine-readable GreenPassport JSON schema mirrors Section~\ref{sec:reporting_standard}. Accounting keys cover the request, service configuration, site, source mix, route, uncertainty, operational predicate, attribution rule, comparator, and output label. Governance keys store the issuer, provenance, validity window, independent verification, and flags. Output labels are \texttt{reject}, \texttt{annual-estimate}, \texttt{scenario}, \texttt{lower-carbon-estimate}, and \texttt{green-eligible}.

\subsection{Request-Energy Attribution}
\label{app:attribution_rule}
Verifiable request energy includes the allocation rule for shared serving energy. GPCA's default measured rule uses separate prefill and decode token-time after subtracting an idle baseline. For serving window $B$, let $E_B$ be measured IT energy, $E_B^{\mathrm{idle}}$ the idle baseline, $\tau^{\mathrm{pre}}_{r,B}$ and $\tau^{\mathrm{dec}}_{r,B}$ request $r$'s accelerator-active prefill and decode token-time, and $M^{\mathrm{kv}}_{r,B}$ its average KV-cache footprint. The default share is
\begin{equation}
\begin{aligned}
a_{r,B}
&=
\frac{\tau^{\mathrm{pre}}_{r,B}+\mu\tau^{\mathrm{dec}}_{r,B}
+\nu M^{\mathrm{kv}}_{r,B}}
{\sum_{r'\in B}\left(
\tau^{\mathrm{pre}}_{r',B}+\mu\tau^{\mathrm{dec}}_{r',B}
+\nu M^{\mathrm{kv}}_{r',B}\right)},\\
E_{\mathrm{it}}(r)
&=\sum_B a_{r,B}\bigl(E_B-E_B^{\mathrm{idle}}\bigr)
+E^{\mathrm{ded}}_r .
\end{aligned}
\nonumber
\end{equation}
Here $\mu$ and $\nu$ are declared weights for decode-time and memory-resident overhead, and $E^{\mathrm{ded}}_r$ stores request-dedicated preprocessing or postprocessing energy. Other valid policies include output-token, GPU-time, batch-slot, marginal-energy, or Shapley-style allocation, and they can assign different values to the same trace. Token-proportional allocation can overcharge long decodes when batching improves utilization, while marginal-energy allocation can be unstable at low load. GPCA makes \texttt{attribution\_rule} a required field.

\subsection{Reporting-Eligibility Algorithm}
\label{app:reporting_algorithm}
Table~\ref{tab:reporting_decision_table} applies the label predicate after the candidate passes memory feasibility and the declared operational predicate.

\begin{table*}[h]
\centering
\caption{GreenPassport label decision rules.}
\label{tab:reporting_decision_table}
\scriptsize
\resizebox{\linewidth}{!}{
\begin{tabular}{@{}L{0.18\linewidth}L{0.24\linewidth}L{0.24\linewidth}L{0.21\linewidth}L{0.20\linewidth}@{}}
\toprule
Label & Required fields & Source requirement & Comparator and uncertainty & Failure mode \\
\midrule
\textsc{reject} & Missing any required request, model/service, site, route, source, document, comparator, or named hardware/instance field & N/A & N/A & Missing input, infeasible hardware, or forbidden data-transfer policy \\
\textsc{annual-estimate} & Complete public entry with annual site or regional CI & Annual public data, public source mix, declared route estimate & Comparator record required. Invalid comparator or $\Delta_c\ge 0$ yields this level & Annual output \\
\textsc{scenario} & Complete scenario entry, including model/hardware defaults, route defaults, and attribution proxy & Public specifications, service disclosure, or declared scenario envelope & Negative point gap with overlapping uncertainty or public inputs & Scenario output \\
\textsc{lower-carbon-estimate} & Complete entry, valid local comparator, attribution rule, and uncertainty interval & Provider disclosure or verified data for the reported boundary & Requires $\sup C_{\mathrm{imp}} < \inf C_{\mathrm{loc}}$ under same-local or best-local comparator & Scenario for interval overlap. Annual estimate for an invalid comparator \\
\textsc{green-eligible} & Complete lower-carbon entry plus clean-energy and verification documents & Hourly or certificate-backed data, deliverability, residual-mix treatment, and no double counting & Robust lower-carbon inequality and valid operational predicate & Recheck eligibility if certificates are double counted or not deliverable \\
\bottomrule
\end{tabular}}
\end{table*}

\noindent\textbf{Pseudocode.}
Given candidate $x$, requested label level $\ell$, and comparator set $\mathcal C$, GPCA constructs $\mathcal R$ and returns \textsc{reject} when $M(x)\neq\emptyset$. It checks memory feasibility and $\Omega(x,c)$ for each eligible comparator $c\in\mathcal C$. An absent comparator record yields \textsc{reject}. A record marked unavailable or invalid yields \textsc{annual-estimate}.

GPCA then computes $C_{\mathrm{imp}}$, $C_{\mathrm{loc}}$, and uncertainty sets from the disclosed energy, route, facility, and carbon-intensity inputs. A nonnegative $\Delta_c$ yields \textsc{annual-estimate}. A negative point gap with public-scenario inputs or overlapping uncertainty sets yields \textsc{scenario}. A robust negative gap under a valid comparator and verified documentation yields \textsc{lower-carbon-estimate}. Hourly or certificate-backed clean-energy data, deliverability, residual-mix treatment, and no double counting enable \textsc{green-eligible}. The procedure returns the highest eligible level.

\noindent\textbf{Strategic-disclosure threat model.}
The verifier checks selective omission of route, PUE, residual mix, or unfavorable operating windows. Further checks cover workload-boundary substitution, allocation changes, stale telemetry or certificates, implausibly low energy relative to throughput, certificate double counting, and unverified provider claims. Missing required fields yield rejection, and cross-field contradictions route the entry to independent review.

\section{Additional Evaluation Details}
\label{app:evaluation_artifacts}

The fixed-workload matrices and metric definitions supplement the main evaluation. Appendix~\ref{app:ablation_study} reports the component ablations.

\subsection{Fixed-Workload Label Matrix}
\begin{figure*}[t]
\centering
\begin{subfigure}[t]{0.48\textwidth}
\centering
\includegraphics[width=\linewidth]{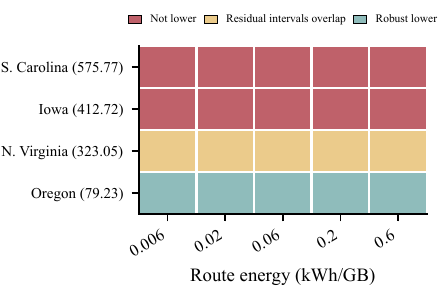}
\caption{GCP U.S. public regions}
\end{subfigure}
\hfill
\begin{subfigure}[t]{0.48\textwidth}
\centering
\includegraphics[width=\linewidth]{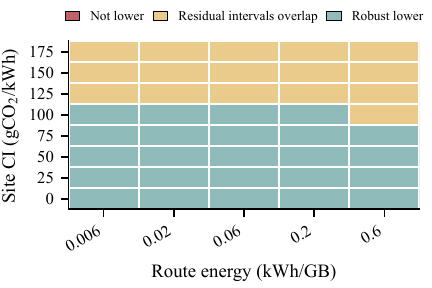}
\caption{CN-West clean scenario}
\end{subfigure}
\caption{Reference-workload label matrix. Panel (a) evaluates four public GCP U.S. regional carbon-intensity rows against the US-Middle comparator. Panel (b) varies the CN-West clean-scenario intensity. Route intensity varies by column while the global sensitivity analysis varies all declared inputs jointly.}
\label{fig:app_b1_global_sensitivity_phase}
\end{figure*}

Fig.~\ref{fig:app_b1_global_sensitivity_phase} shows fixed-workload decision matrices for public GCP U.S. regions and the CN-West clean scenario. The main sensitivity analysis varies all declared inputs jointly.

\subsection{Tool Boundary Comparison}
Appendix~Table~\ref{tab:tool_boundary} compares the accounting inputs and reporting components of the tools discussed in the main evaluation.

\begin{table*}[t]
\centering
\caption{Tool boundary comparison.}
\label{tab:tool_boundary}
\scriptsize
\begin{tabular}{@{}lccccccc@{}}
\toprule
Method & Local energy & Remote API & Site CI/PUE & Network & Request attribution & Uncertainty & Reporting basis \\
\midrule
CodeCarbon & Measure/estimate & -- & Yes & -- & External & -- & -- \\
CarbonTracker & Measure/predict & -- & Yes & -- & External & Forecast & -- \\
EcoLogits & -- & Estimate & Region/model & Partial & Request & Range & -- \\
GPCA & Import/estimate & Passport & Yes & Bounded & Declared rule & Typed sets & Gate/provenance \\
\bottomrule
\end{tabular}
\end{table*}
 
\subsection{Evaluation Metrics}
\label{app:evaluation_metrics}

Request carbon combines serving-site carbon and route carbon.
\[
C_{\mathrm{req}}=C_{\mathrm{site}}+C_{\mathrm{route}},\qquad [\text{\gco/request}] .
\]
Token carbon normalizes request carbon by output length.
\[
C_{\mathrm{tok}}^{\mathrm{mg}}=1000 C_{\mathrm{req}}/t^{out},\qquad [\text{mgCO$_2$e/token}] .
\]
Percentage reductions use the same-local and best-local comparators. Positive values indicate lower carbon.
\[
\begin{aligned}
R_{\mathrm{same}}
&=100\,
\frac{C_{\mathrm{req}}(c_{\mathrm{same}})-C_{\mathrm{req}}(x)}
{C_{\mathrm{req}}(c_{\mathrm{same}})},\\
R_{\mathrm{best}}
&=100\,
\frac{C_{\mathrm{req}}(c_{\mathrm{best}})-C_{\mathrm{req}}(x)}
{C_{\mathrm{req}}(c_{\mathrm{best}})} .
\end{aligned}
\]
Selection regret measures the carbon penalty of the candidate selected by a baseline.
\[
\mathrm{Regret}=
\frac{C_{\mathrm{req}}(x_{\mathrm{selected}})-C_{\mathrm{req}}(x_{\mathrm{oracle}})}
{C_{\mathrm{req}}(x_{\mathrm{oracle}})},
\]
where $x_{\mathrm{oracle}}$ is the lowest-carbon feasible and admissible candidate under GPCA. Route contribution is
\[
\rho_{\mathrm{route}}=C_{\mathrm{route}}/C_{\mathrm{req}} .
\]
Declared method-schema completeness, reporting eligibility, and rule-overstatement rates are computed over the deterministic fixtures. The complementary internal conformance score is
\[
\mathrm{RuleConsistency}=1-\mathrm{RuleOverstatementRate},
\]
where a rule-consistent output stays within the level supported by its inputs, comparator, uncertainty, route boundary, feasibility, and attribution rule. The deterministic property test follows Table~\ref{tab:reporting_decision_table}. Omitted components remain missing inputs in the ablations.

The reporting-rule fixtures span annual estimates, lower-carbon estimates against same-local and best-local comparators, and green-eligible labels. A method overstates when its requested label exceeds the level returned by the executable gate in Table~\ref{tab:reporting_decision_table}. The generator instantiates a \texttt{GreenPassport}, calls \texttt{admissible\_output\_label} for every case, writes requested and returned labels to the case-level CSV, validates denominators, and recomputes component-omission regret. These deterministic fixtures measure implementation conformance to the decision table.

\subsection{Case-Study Conditional Decision Card}
Table~\ref{tab:eval_recommendation_card} gives the US-Middle reference-point result and its stability across the uncertainty envelope.

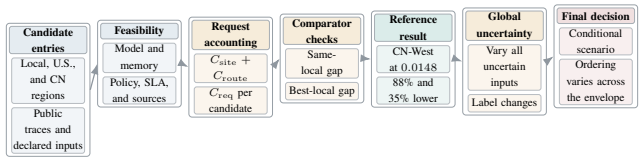
\begin{figure}[t]
\centering
\captionsetup{type=figure,hypcap=false}
\resizebox{\columnwidth}{!}{%
\begin{tikzpicture}[
    node distance=0.08cm,
    stagehead/.style={draw=BAGTAGray!55, rounded corners=1.5pt, align=center, text width=1.62cm, minimum height=0.34cm, inner sep=1pt, font=\scriptsize\bfseries},
    subblock/.style={draw=BAGTAGray!45, rounded corners=1.5pt, align=center, text width=1.52cm, minimum height=0.38cm, inner sep=1pt, font=\scriptsize},
    stage/.style={draw=BAGTAGray!65, rounded corners=2pt, inner sep=0.06cm},
    arr/.style={-Latex, thick, draw=BAGTAGray!70}
]
\node[stagehead, fill=BAGTABlue!12] (ah) {Candidate entries};
\node[subblock, fill=BAGTABlue!6, below=of ah] (a1) {Local, U.S., and CN regions};
\node[subblock, fill=BAGTABlue!6, below=of a1] (a2) {Public traces and declared inputs};
\node[stage, fit=(ah)(a1)(a2)] (a) {};

\node[stagehead, fill=BAGTABlue!12, anchor=north west] (bh) at ([xshift=0.22cm]a.north east) {Feasibility};
\node[subblock, fill=BAGTABlue!6, below=of bh] (b1) {Model and memory};
\node[subblock, fill=BAGTABlue!6, below=of b1] (b2) {Policy, SLA, and sources};
\node[stage, fit=(bh)(b1)(b2)] (b) {};

\node[stagehead, fill=BAGTAGold!16, anchor=north west] (ch) at ([xshift=0.22cm]b.north east) {Request accounting};
\node[subblock, fill=BAGTAGold!8, below=of ch] (c1) {$C_{\mathrm{site}}+C_{\mathrm{route}}$};
\node[subblock, fill=BAGTAGold!8, below=of c1] (c2) {$C_{\mathrm{req}}$ per candidate};
\node[stage, fit=(ch)(c1)(c2)] (c) {};

\node[stagehead, fill=BAGTAGold!16, anchor=north west] (dh) at ([xshift=0.22cm]c.north east) {Comparator checks};
\node[subblock, fill=BAGTAGold!8, below=of dh] (d1) {Same-local gap};
\node[subblock, fill=BAGTAGold!8, below=of d1] (d2) {Best-local gap};
\node[stage, fit=(dh)(d1)(d2)] (d) {};

\node[stagehead, fill=BAGTATeal!16, anchor=north west] (eh) at ([xshift=0.22cm]d.north east) {Reference result};
\node[subblock, fill=BAGTATeal!8, below=of eh] (e1) {CN-West at $0.0148$};
\node[subblock, fill=BAGTATeal!8, below=of e1] (e2) {88\% and 35\% lower};
\node[stage, fit=(eh)(e1)(e2)] (e) {};

\node[stagehead, fill=BAGTAGold!16, anchor=north west] (fh) at ([xshift=0.22cm]e.north east) {Global uncertainty};
\node[subblock, fill=BAGTAGold!8, below=of fh] (f1) {Vary all uncertain inputs};
\node[subblock, fill=BAGTAGold!8, below=of f1] (f2) {Label changes};
\node[stage, fit=(fh)(f1)(f2)] (f) {};

\node[stagehead, fill=BAGTARed!16, anchor=north west] (gh) at ([xshift=0.22cm]f.north east) {Final decision};
\node[subblock, fill=BAGTARed!8, below=of gh] (g1) {Conditional scenario};
\node[subblock, fill=BAGTARed!8, below=of g1] (g2) {Ordering varies across the envelope};
\node[stage, fit=(gh)(g1)(g2)] (g) {};

\foreach \u/\v in {a/b,b/c,c/d,d/e,e/f,f/g} {\draw[arr] (\u.east) -- (\v.west);}
\end{tikzpicture}
}
\caption{Buyer decision chain from the reference point to the uncertainty-envelope result.}
\label{fig:eval10_buyer_funnel}
\end{figure}

\begin{table}[h]
\centering
\caption{US-Middle decision card.}
\label{tab:eval_recommendation_card}
\small
\resizebox{\linewidth}{!}{
\begin{tabular}{@{}ll@{}}
\toprule
Field & Reference-point result \\
\midrule
Buyer region & US-Middle \\
Point-estimate serving region & CN-West scenario \\
Model class & Medium assistant model \\
Accelerator/instance & B200-class, feasible high-memory setting \\
Request carbon & $0.0148$ \gco/request \\
Site carbon & $0.0144$ \gco/request \\
Route carbon & $0.00038$ \gco/request \\
Token carbon & $0.030$ mgCO$_2$e/token \\
Site carbon share & $97.4\%$ \\
Route carbon share & $2.6\%$ \\
Comparator & GCP Oregon best-local and US-Middle same-local \\
Output label & \textsc{scenario} \\
Required inputs & Same-hour clean-site and route data \\
Global stability & Ordering varies across the input ranges \\
\bottomrule
\end{tabular}
}
\end{table}

\setcounter{section}{3}
\setcounter{figure}{0}
\begin{figure*}[tp]
\centering
\captionsetup{skip=2pt}
\begin{minipage}[t]{0.46\textwidth}
\centering
\includegraphics[width=\linewidth]{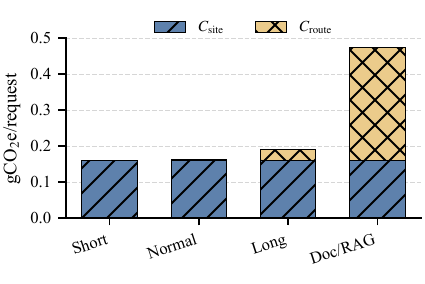}
\caption{Route and payload sensitivity ($\downarrow$).}
\label{fig:app_c1_route_payload}
\end{minipage}\hfill
\begin{minipage}[t]{0.46\textwidth}
\centering
\includegraphics[width=\linewidth]{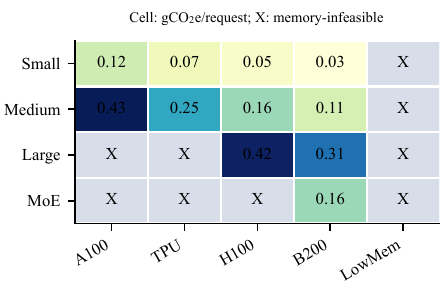}
\caption{Accelerator and instance sensitivity ($\downarrow$).}
\label{fig:app_c2_accel_instance}
\end{minipage}
\end{figure*}

\begin{figure*}[t]
\centering
\captionsetup{skip=2pt}
\begin{minipage}[b]{0.6\linewidth}
    \centering
\begin{subfigure}[t]{0.47\linewidth}
\centering
\includegraphics[width=\linewidth]{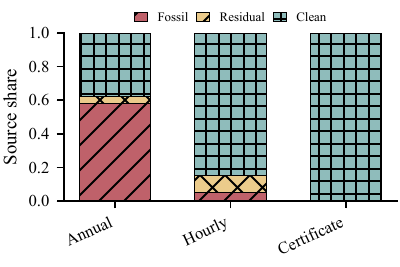}
\caption{Source basis}
\end{subfigure}\hfill
\begin{subfigure}[t]{0.47\textwidth}
\centering
\includegraphics[width=\linewidth]{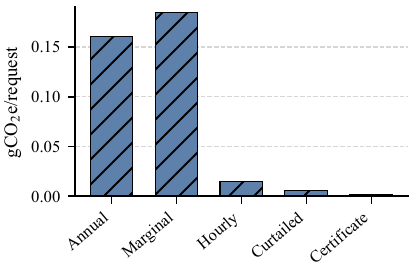}
\caption{Time basis ($\downarrow$)}
\end{subfigure}
\caption{Source and time-basis sensitivity.}
\label{fig:app_c3_source_time}
\end{minipage}\hfill
\begin{minipage}[b]{0.4\linewidth}
\centering
\vspace{1em}
\includegraphics[width=0.9\linewidth]{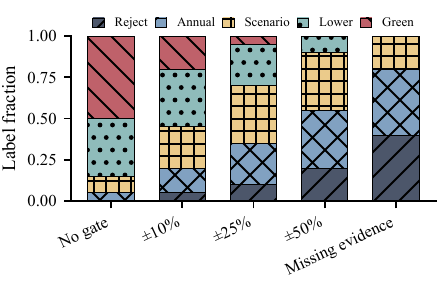}
\caption{Label transition under uncertainty and source-quality gates.}
\label{fig:app_c4_label_uncertainty}
\end{minipage}
\end{figure*}

\begin{figure*}[!tp]
\centering
\begin{subfigure}[t]{0.24\linewidth}
\centering
\includegraphics[width=\linewidth]{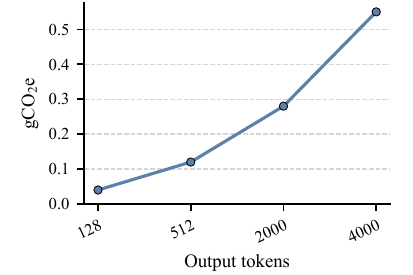}
\caption{$t^{out}$}
\end{subfigure}
\hfill
\begin{subfigure}[t]{0.24\linewidth}
\centering
\includegraphics[width=\linewidth]{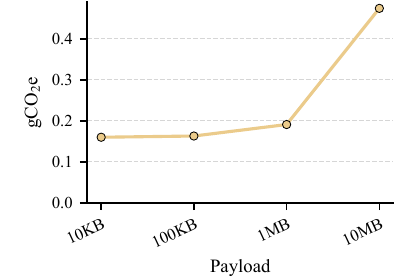}
\caption{$B_r$}
\end{subfigure}
\hfill
\begin{subfigure}[t]{0.24\linewidth}
\centering
\includegraphics[width=\linewidth]{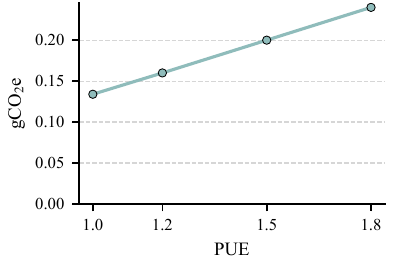}
\caption{$\phi_k$}
\end{subfigure}
\hfill
\begin{subfigure}[t]{0.24\linewidth}
\centering
\includegraphics[width=\linewidth]{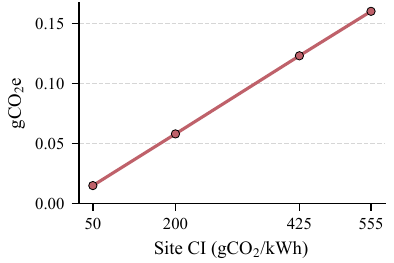}
\caption{$I_s$}
\end{subfigure}
\caption{Controlled sensitivity ($\downarrow$). Output length, payload, PUE, and serving-site carbon intensity move request carbon monotonically.}
\label{fig:app_c5_controlled_sensitivity}
\end{figure*}

\setcounter{section}{2}
\section{Ablation Study}
\label{app:ablation_study}

The ablation removes one GPCA mechanism at a time while retaining the workload data, candidate set, public inputs, and carbon coefficients. It measures changes in request carbon, candidate feasibility, comparator outcome, and admissible label.

\begin{table*}[t]
\centering
\small
\begin{tabular}{p{0.16\textwidth}p{0.43\textwidth}p{0.31\textwidth}}
\toprule
Variant & Change from Full GPCA & Isolated effect \\
\midrule
NoRoute & Remove delivery-route carbon from every candidate & Importance of the cross-border delivery boundary \\
NoInstanceGate & Admit candidates without the memory-feasibility check & Effect of hardware and instance feasibility \\
SameLocalOnly & Replace the best feasible local comparator with the same local configuration & Effect of comparator choice on the reported reduction \\
PointEstimate & Apply the label decision to point estimates without the combined uncertainty envelope & Effect of uncertainty on comparative stability \\
NoSourceGate & Assign labels without source, time-basis, and reporting-tier requirements & Effect of source quality on label strength \\
\bottomrule
\end{tabular}
\caption{GPCA component ablations.}
\label{tab:gpca_organized_ablations}
\end{table*}

Section~\ref{subsec:interpretation} reports the main component effects. Fig.~\ref{fig:app_c1_route_payload} breaks down route carbon by workload, and Fig.~\ref{fig:app_c2_accel_instance} shows carbon and memory feasibility across the accelerator catalog. Figs.~\ref{fig:eval6_comparator_reduction} and~\ref{fig:eval7_selection_regret} compare local references and component-omission regret.

Annual, marginal, hourly, curtailed, and certificate-backed inputs produced different estimates and output levels in Fig.~\ref{fig:app_c3_source_time}. Fig.~\ref{fig:app_c4_label_uncertainty} gives the label transitions as uncertainty and source requirements changed. With the full method active, request carbon increased monotonically with output length, payload, PUE, and serving-site carbon intensity in Fig.~\ref{fig:app_c5_controlled_sensitivity}.

\section{Serving Implementation Context}

MLPerf Inference provides the corresponding performance and configuration context \cite{mlperf51}. Facility studies characterize data-center overhead and electricity demand \cite{masanet2020recalibrating,iea2023datacentres,iea2025energyai}.

FlashAttention changes attention execution \cite{dao2022flashattention,dao2024flashattention2}, speculative decoding changes the number of target-model steps \cite{leviathan2023speculative}, and MoE routing and quantization change active computation and memory traffic \cite{lepikhin2021gshard,dettmers2023qlora}.

\section*{Scope}

The evaluation used public ML.ENERGY aggregate configurations for the represented workloads and H100- and B200-class accelerators. Route-class intensity intervals and annual, regional, or scenario electricity data supplied the remaining inputs. Higher reporting levels require provider telemetry, time-matched electricity data, route measurements, and the corresponding verification documents.

The accounting boundary includes operational carbon from serving and delivery. It excludes training amortization, embodied hardware emissions, water, land, price, reliability, privacy, and grid-expansion effects.
 
\end{document}